\documentclass[fleqn,usenatbib]{mnras}

\usepackage{newtxtext,newtxmath}

\usepackage[T1]{fontenc}
\usepackage{ae,aecompl}

\usepackage{graphicx}	
\usepackage{amsmath}	
\usepackage{amssymb}	

\usepackage[export]{adjustbox}

\usepackage{xcolor}

\usepackage{ulem}

\defcitealias{Ormel2015I}{Paper I}
\defcitealias{Ormel2015II}{Paper II}

\definecolor{mygray}{gray}{0.6}

\newcommand{\coadd}[1]{\textcolor{blue}{{#1}}}

\newcommand{\minc}[1]{{{#1}}}

\title[Hydrodynamics of embedded planets' first atmospheres - III]{Hydrodynamics of embedded planets' first atmospheres - III. The role of radiation transport for super-Earth planets}

\author[N. P. Cimerman et al.]{
Nicolas P. Cimerman,$^{1}$\thanks{E-mail: npcphys@gmail.com (NPC)}
Rolf Kuiper,$^{1}$
Chris W. Ormel$^{2}$
\\
$^{1}$Institut f\"ur Astronomie \& Astrophysik, Universit\"at T\"ubingen, Auf der Morgenstelle 10, D-72076 T\"ubingen, Germany\\
$^{2}$Anton Pannekoek Institute, University of Amsterdam, PO Box 94249, NL-1090 GE Amsterdam, the Netherlands
}

\date{Accepted 2017 July 22. Received 2017 July 18; in original form 2017 June 7}

\pubyear{2017}

\begin{document}
\label{firstpage}
\pagerange{\pageref{firstpage}--\pageref{lastpage}}
\maketitle

\begin{abstract}
The population of close-in \textit{super-Earths}, with gas mass fractions of up to
10\% represents a challenge for planet formation theory: how did they avoid runaway gas accretion
and collapsing to hot Jupiters despite their core masses being in the critical range of
$M_{\mathrm{c}} \simeq 10 \, M_{\oplus}$?
Previous three-dimensional (3D) hydrodynamical simulations indicate that atmospheres of low-mass planets cannot be considered isolated from the protoplanetary disc, contrary to what is assumed in 1D-evolutionary calculations. This finding is referred to as the \textit{recycling} hypothesis.
In this Paper we investigate the recycling hypothesis for super-Earth planets, accounting for realistic 3D radiation-hydrodynamics. Also, we conduct a direct comparison in terms of the evolution of the entropy between 1D and 3D geometries. We clearly see
that 3D atmospheres maintain higher entropy: although gas in the atmosphere loses \textit{entropy} through radiative cooling, the advection of high entropy gas from the disc into the Bondi/Hill sphere slows down Kelvin-Helmholtz contraction, potentially arresting envelope growth at a sub-critical gas mass fraction.
Recycling, therefore, operates vigorously, in line with results by previous studies. However, we also identify an ``inner core'' -- in size $\approx$\,$25$\% of the Bondi radius -- where streamlines are more circular and entropies are much lower than in the outer atmosphere. Future studies at higher resolutions are needed to assess whether this region can become hydrodynamically-isolated on long time-scales.
\end{abstract}




\section{Introduction}

Giant planets form in gas rich discs either by gravitational instability \citep{Kuiper1951,Cameron1978} or, as is assumed more commonly, by core accretion \citep{Safronov1969,PollackEtal1994}. In the latter scenario the final rapid gas accretion proceeds only after a massive enough `core' has been assembled. 
A typically quoted number for this critical core mass (CCM) is $\sim$10 Earth mass \citep{Stevenson1982,PollackEtal1994} -- a number that follows from solving 1D stellar-structure like equations. Assuming radial symmetry, these models compute how the temperature, pressure, and density profiles evolve as function of time. As pointed out by \citet{PisoYoudin2014}, these models come in different flavours: static models consider accretion of solids to provide a steady substantial luminosity to support the envelope \citep[e.g.][]{Stevenson1982,Rafikov2006}, while in quasi-static models hydrostatic equilibrium is assumed at any given time, but the Kelvin-Helmholtz contraction of the atmosphere is considered as source of the luminosity \citep[e.g.][]{IkomaEtal2001,LeeChiang2015}.


The $\sim$10 $M_\oplus$ for the CCM is typical for the formation of Jupiter assuming ISM-like grain opacities (in regions cool enough for silicate grains to be present), a steady influx of planetsimals, and a H/He-dominated envelope (all solids end up in the core) \citep{PollackEtal1994,HubickyjEtal2005,MordasiniEtal2009}. However, in the last years it has been realized that some of these model parameters are chosen rather conservatively. In envelopes, grains will quickly coagulate and settle \citep{MovshovitzEtal2010,Mordasini2014,Ormel2014}. A zero grain opacity may be more realistic, especially in the case without solid accretion. It is also rather restrictive to assume continuous planetesimal accretion. Without an external source for luminosity and grain opacity \citet{HoriIkoma2010} calculate that cores as low as $2\,M_\oplus$ will collapse into a gas giant after 1 Myr. A metal-rich composition of the envelope will also reduce the CCM \citep{VenturiniEtal2015}.

These considerations are especially important for super-Earth planets. Super-Earths are a class of planet that lie by mass in between terrestrial (`rocky') planets and giant planets. They have been found in great numbers by transit and radial-velocity surveys, in particular by the \textit{Kepler} spacecraft \citep{FressinEtal2013,PetiguraEtal2013}, at short distances (typically $\sim$0.1\,au) from their host star. With radial-velocity (RV) follow-up \citep{MayorEtal2011,MarcyEtal2014} or transit timing variations (TTV) \citep{HaddenLithwick2014,Jontof-Hutter2016,MillsMazeh2017}
a bulk density can be obtained, which, for the more massive super-Earths, is low enough to infer that these planets contain significant amounts ($\sim$1\% by mass) of hydrogen and helium \citep{LopezFortney2014}.
These super-Earths are also referred to as mini-Neptunes. With these hydrogen/helium mass fractions, it is plausible to assume that these planets formed early, in a gas-rich disc, like giant planets do. Even the current rocky super-Earths (\textit{i.e.}, those planets of high bulk density) could have started with gas-rich envelopes, only to have their atmospheres stripped by photo-evaporation \citep{LopezFortney2013,LopezFortney2014,OwenWu2013} or lost in a ``boil-off'' event, when the confining pressure of the disc is gone after dispersal \citep{OwenWu2016}.
A further analysis based on more precise radii from the \textit{California-Kepler Survey} by \citet{FultonEtal2017} provides some evidence for such a scenario which has been recently
discussed by \citet{OwenWu2017} and \citet{JinMordasini2017}.

When super-Earths did assemble before the disc dispersal, the question is how they avoided breaching the CCM, which would have turned them into hot-Jupiters. A key point is that at $\sim$0.1\,au planetesimals or any other solid particles \coadd{are} accreted on time-scales $\ll$\,Myr \citep{ChiangLaughlin2013}. Modelling super-Earths envelopes including non-ideal effects (like hydrogen dissociation and silicates evaporation), \citet{Lee2014} calculated that the Kelvin-Helmholtz contraction took $\sim$Myr. To avoid envelope collapse \citet{LeeChiang2015} suggested super-Earths resided in gas-depleted transition discs. The little time that is left for the planet to accrete remnant gas during the disk dispersal phase is limiting in this scenario, rather than the reduced gas density. In such a disc, super-Earths could form from a bunch of planetary embryos, similar to well-established models for the inner solar system \citep{DawsonEtal2016}. Other works have continued to search for a thermodynamic explanation to prolong the cooling time of super-Earth planets \citep{BatyginEtal2016,ColemanEtal2017}. When super-Earths are on eccentric orbits ($e\sim0.1$) \citet{GinzburgSari2017} showed that stellar tidal heating would suppress the cooling.

In contrast, we favour a hydrodynamical explanation. It was shown in \citet{Ormel2015II} (hereafter referred to as \citetalias{Ormel2015II}) that envelopes are not isolated from the disc; instead, they exchange gas with the disc. Unlike the 2D case (\citealt{Ormel2015I}; hereafter \citetalias{Ormel2015I}), where we saw closed streamlines around the planet, in 3D horseshoe and in-spiralling streamlines effectively recycle material between the disc and `envelope' (in an open system the very definition of 'envelope' becomes arbitrary). Similar results were found by \citet{WangEtal2014} and \citet{FungEtal2015}. Importantly, atmosphere recycling will arrest thermodynamical cooling when the atmosphere recycling timescale (the time it takes to replenish the planets' atmosphere; \citetalias{Ormel2015II}) is shorter than the Kelvin-Holmholtz contraction time-scale. In \citetalias{Ormel2015II} we argued that recycling should be particularly effective for super-Earths: they have rather modest gravitational potentials (because of their low ratio of physical to Hill radii) and due to their large orbital speeds they encounter much disc gas per unit time.
    
However, the simulations of \citetalias{Ormel2015II} were conducted for a Bondi radius-to-gas scale-height of 1\%,\footnote{Throughout this work, we define the Bondi radius as $R_\mathrm{Bondi}=GM/c_\mathrm{s}^2$, where $c_\mathrm{s}$ is the (isothermal) sound speed of the background disc gas. Also, we take the gas scale-height as $H=c_\mathrm{s}/\Omega$ where $\Omega$ is the local orbital frequency, and the Hill radius is $R_\mathrm{Hill}=r (M_\mathrm{p}/3M_\star)^{(1/3)}$ where $r$ is the disc orbital radius, $M_\mathrm{p}$ the planet mass and $M_\star$ the stellar mass. With these definitions we have the relation $R_\mathrm{H}^3=R_\mathrm{B} H^2/3$. For the close-in super-Earth planets that we consider, usually we have $R_\mathrm{B} \simeq R_\mathrm{H}$.} which is representative of a Mars-sized embryo at 1\,au, not for super-Earth planets. In addition, the \citetalias{Ormel2015II} simulations considered an isothermal equation of state and, because of the computationally-intensive nature of the simulations, could only run the simulation for several dynamical times. Here, to test whether the recycling hypothesis is truly applicable for super-Earths, we conduct simulations that (i) involve super-Earth planets (up to $5\,M_\oplus$ cores at $0.1\,$au); (ii) utilize proper radiation transport to consider the cooling of their atmospheres; and (iii) run the simulations for much longer times.

With these improved physics, our key goal is to test the recycling hypothesis for super-Earth planets. We do observe a very similar flow topology as in \citetalias{Ormel2015II}, with disc streamlines penetrating deep in the Bondi sphere. However, unlike \citetalias{Ormel2015II}, we also see a low-entropy ``inner core'' emerging where streamlines are more circular. Indeed, the recycling hypothesis can best be tested by following the entropy. Radiative cooling decreases entropy and in every 1D thermal evolutionary model entropy always decreases with time (since the planet is hydrodynamically isolated). However, recycling has the opposite effect: high entropy gas from the disc enters the envelope, whereas low-entropy gas flows back to the disc. To quantitatively test the influence of recycling we conduct a direct comparison of radiative hydrodynamical simulations between 1D and 3D geometries. Although the 3D simulations can only run up to $\sim$$10^2$--$10^3$ orbits, this comparison unambiguously shows that recycling delays the thermal cooling of super-Earths' atmospheres. On the other hand, the evolution time is usually not sufficient to evolve simulations into a steady state ($t \ngg$ recycling time of the interior atmosphere), where recycling would finally arrest radiative cooling.

The paper is organized as follows: In Section 2 we discuss the physical setup of our model, highlighting differences to \citetalias{Ormel2015I} and \citetalias{Ormel2015II}. Initial conditions are explained in Section 3. In Section 4 we describe the numerical configuration. Results are presented in Section 5 and discussed in Section 6. To conclude, a summary is made and followed by a brief outlook in Section 7.

\section{Theory and Numerics}
In this section we first discuss the equations of Radiation-Hydrodynamics in the Flux-Limited-Diffusion (FLD) approximation for radiation transfer \citep{Levermore1981}.
Secondly, we explain the forces that operate in the adopted local frame co-orbiting
with the planet.
Then we describe the numerical methods used to solve the governing equations.

\subsection{Equations of Radiation-Hydrodynamics}
Following previous studies \citepalias{Ormel2015I, Ormel2015II} we describe the disc gas as an inviscid\footnote{We note that in numerical models, there is always some intrinsic numerical viscosity. Using sufficiently high resolution, this effect is usually negligible. In Appendix A of \citetalias{Ormel2015I}, this issue is discussed in more detail using a similar geometry and the same \texttt{roe} solver.} and compressible fluid.
In this study, we drop the previously made isothermal assumption and solve for the radiative transport of energy.
We employ an ideal equation of state:
\begin{align}
	P = \left(\gamma - 1 \right)\rho e_{\mathrm{gas}}
	\equiv \exp (s) \,\rho^{\gamma},
	\label{eq:eos}
\end{align}
with the dimensionless entropy $s = S/c_\mathrm{V}$, where $S$ is entropy, $c_\mathrm{V}$ the heat capacity
at constant volume and $\gamma$ is the adiabatic exponent.
We consider local thermal equilibrium (LTE) and use the one-temperature approximation,
assuming the gas and radiation to have the same temperature
$T_{\mathrm{gas}} = T_{\mathrm{rad}} = T$.
The three basic equations that account for conservation of mass, momentum and energy for
such a fluid are the continuity, Euler's and the energy equation:
\begin{align}
	\frac{\partial \rho}{\partial t} + \boldsymbol{\nabla} \cdot \left( \rho \mathbfit{v} \right) &= 0 \\
	\left( \frac{\partial }{\partial t} + \mathbfit{v} \cdot \boldsymbol{\nabla} \right) \mathbfit{v}
	&= -\frac{\boldsymbol{\nabla} P_\mathrm{gas}}{\rho} + \sum\limits_{i} \mathbfit{a}_i,\\
	\frac{\partial E}{\partial t} + \boldsymbol{\nabla} \cdot \left(E \mathbfit{v} \right)
	&= - \boldsymbol{\nabla} \cdot \mathbfit{F} - P_\mathrm{gas} \boldsymbol{\nabla} \cdot \mathbfit{v}
	\label{eq:etot}
\end{align}
where $\rho$ is density, $t$ time, $P_\mathrm{gas}$ gas pressure, $\mathbfit{v}$ the gas velocity
$E = E_{\mathrm{int}} + E_{\mathrm{rad}}$ the sum of internal and radiation energy and
$\mathbfit{F}$ the flux of radiation energy density. Accelerations due to external forces are
represented by the $\mathbfit{a}_i$.
Decomposing the energy equation into a component for internal and radiation energy, we write the internal energy density as
$E_{\mathrm{int}} = c_{V} \rho T$,
and by virtue of the assumed LTE, the radiation energy density is given by
$E_{\mathrm{rad}} = a T^4$, where $a$ is the radiation constant.
Their time evolution is governed by the following equations
\citep[e.g.][]{Kuiper2010}:
\begin{align}
	\frac{\partial{E_{\mathrm{int}}}}{\partial t} + \boldsymbol{\nabla}
	\cdot \left( E_{\mathrm{int}} \mathbfit{v}
	\right) &= -P_\mathrm{gas} \boldsymbol{\nabla} \cdot \mathbfit{v} + \Lambda
	\label{eq:eint}\\
	\frac{\partial{E_{\mathrm{rad}}}}{\partial t} + \boldsymbol{\nabla}
	\cdot \left( E_{\mathrm{rad}} \mathbfit{v}
	\right) &= - \boldsymbol{\nabla} \cdot \mathbfit{F} - \Lambda,
	\label{eq:erad}
\end{align}
where $\Lambda$ describes the coupling term of matter and radiation, which in LTE
vanishes.
The (FLD) flux of radiation energy density is then given by
\begin{align}
	\mathbfit{F} = \frac{\lambda c}{\rho \kappa_{\mathrm{R}}} \boldsymbol{\nabla} E_{\mathrm{rad}},
\end{align}
with the flux-limiter $\lambda$, the speed of light $c$ and the Rosseland mean opacity
$\kappa_{\mathrm{R}}$. For the choice of the flux-limiter, we follow \citet{Levermore1981},
ensuring that the flux of radiation energy is described correctly in the limiting cases
of free streaming in the optically thin regime and pure diffusion in optically thick regions.

Using operator splitting we solve for the transport term $\boldsymbol{\nabla} \cdot
\left( E \mathbfit{v} \right)$ separately during the hydrodynamical step so that the equation to be
solved in the radiation/FLD step reads
\begin{align}
	\frac{\partial E}{\partial t} 
	&= - \boldsymbol{\nabla} \cdot \mathbfit{F} - P_\mathrm{gas} \boldsymbol{\nabla} \cdot \mathbfit{v} .
	\label{eq:E_FLD}
\end{align}
Using the relation
\begin{align}
	\frac{\partial E_{\mathrm{int}}}{\partial t} = \frac{c_{\mathrm{V}} \rho}{4 a T^3} 
	\frac{\partial E_{\mathrm{rad}}}{\partial t},
\end{align}
we can rewrite Eqn. (\ref{eq:E_FLD}) as a diffusion equation
\begin{align}
	\frac{\partial E_{\mathrm{rad}}}{\partial t} =
    \frac{- \boldsymbol{\nabla} \cdot \mathbfit{F} - P_\mathrm{gas} \boldsymbol{\nabla} \cdot \mathbfit{v}}{1 + \frac{c_{\mathrm{V}} \rho}{4 a T^3}}.
    \label{eq:fldeq}
\end{align}
For more technical details, see e.g. \citet{Kuiper2010}.

In the temperature regime that we find in our simulations
($ 3 \lesssim \log T[\mathrm{K}] \lesssim 4$), the pressure due to radiative forces
${P_{\mathrm{rad}} = E_{\mathrm{rad}} / 3 \ll P_{\mathrm{gas}} = \rho k_{\mathrm{B}} T /
\mu m_{\mathrm{H}}}$ is small
compared to the gas pressure. Here, $k_{\mathrm{B}}$ is Boltzmann's constant and
$m_{\mathrm{H}}$ is the mass of atomic hydrogen. In this regime, the total energy budget is dominated by the internal gas energy
($E_{\mathrm{rad}} \ll E_{\mathrm{int}}$) and the transport of energy will be dominated by the flux of radiation energy $\mathbfit{F}$.
We neglect forces due to radiation and
do not account for ionization or dissociation of molecules, taking a constant mean molecular
weight $\mu = 2.353$, corresponding to gas of solar metallicity.
Since molecular hydrogen is the main constituent, the ratio of heat capacities or
adiabatic index is taken as ${\gamma = 7/5}$.
We note that most one-dimensional
\citep[e.g.][]{Lee2014} and some three-dimensional studies \citep[e.g.][]{DAngelo2013} include
much more complex thermodynamics, for example by accounting for ionization and dissociation of
molecules in their opacities and equation of state. However, since we are primarily interested in
the effects of the flow structure, we chose a simple model.

\subsection{Forces in the local frame of reference}
For all our simulations we assume the planet to be on a circular orbit with a fixed separation
$a$ from the host star.
We then adopt a local non-inertial frame of reference that is centred on the planet
and follows its motion around the star. Given that the planet's mass for
a low-mass gaseous envelope is determined by the core mass $M_{\mathrm{p}}
\simeq M_{\mathrm{c}}$, and
the stellar mass that we set as $M_{\star} = 1 \, M_{\sun}$, the Keplerian angular velocity of the planetary
core's orbit is
\begin{align}
	\Omega = \sqrt{\frac{G (M_{\sun} + M_{\mathrm{c}})}{a^3}} \simeq
	\sqrt{\frac{G M_{\sun}}{a^3}}.
\end{align}

\begin{figure*}
	\centering
	\includegraphics[width=17cm]{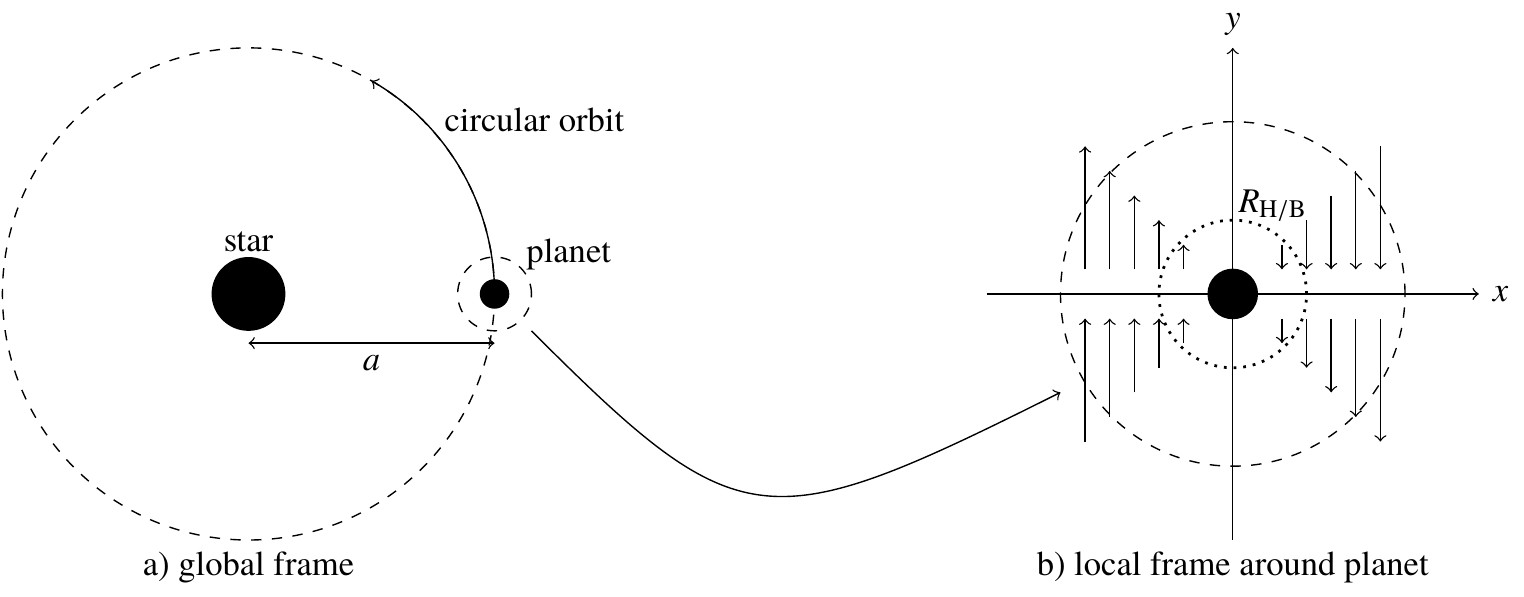}
	\caption{Schematic drawing of local shearing sheet approximation. a) The planet is fixed on a circular orbit with the seperation $a$ from
	the star. b) We align our co-moving frame that is centred on the planet such that the $x$-direction is pointing radially away from
	the star and the $y$-direction follows the orbital direction of the planet. Arrows indicate the gas velocity of the shearing flow.
    Without a ``headwind'' arising from a pressure gradient in the disc, this flow is symmetric with respect to the planet.}
	\label{fig:shearsheet}
\end{figure*}

We set up our frame to be centred on the planet, with the $x$-direction
pointing radially away from the star and the $y$-direction following the orbital direction
of the planet's orbit. This ``shearing-box'' approach is a standard tool to study effects in discs on a smaller, local scale \citep[e.g.][]{GoldreichLB1965,GoldreichTremaine1978,Hawley1995}.
In Fig. \ref{fig:shearsheet} we show a schematic drawing of the setup.
The axis of rotation is aligned with the direction of the $z$-axis, so that we can
write the local Keplerian frequency as $\mathbf\Omega = \Omega \,\mathbfit{e}_z$. Since our
local frame is non-inertial, we have to consider accelerations due to the Coriolis force,
${\mathbfit{a}_{\mathrm{cor}} = -2 \,\mathbf\Omega \times \mathbfit{v}}$\footnote{In \citetalias{Ormel2015I} there is a missing minus sign in the definition of the Coriolis force.} and the centrifugal force
${\mathbfit{a}_{\mathrm{cent}} = -\mathbf\Omega \times \left( \mathbf\Omega \times \mathbfit{r} \right)}$. In a linear
approximation (for $x,y,z \ll a$), the latter can be put together with contributions of stellar
gravity to give the tidal acceleration
${\mathbfit{a}_{\mathrm{tid}} = 3 x \,\Omega^2 \,\mathbfit{e}_x - z\, \Omega^2 \,\mathbfit{e}_z}$.
The vertical component of $\mathbfit{a}_{\mathrm{tid}}$ gives rise to a density stratification of the disc in $z$ and was not considered in the low-mass limit \citepalias{Ormel2015I, Ormel2015II}, because there $R_\mathrm{Bondi}/H\ll1$.
Super-Earths, however, are more massive planets, so that this additional term cannot be neglected.

We only consider a Keplerian disc, so we do not take into account the global pressure force,
which results in a sub-Keplerian motion of the gas, seen as a ``headwind'' in the planet's frame,
as described in \citetalias{Ormel2015I, Ormel2015II}. Lastly, we have to take care of the gravitational
interaction of the disc gas with the planetary core.
As in \citetalias{Ormel2015II}, we use an exponential
smoothing term in space to modify the Newtonian force in a way that results in a force-free
boundary at the core's surface:
\begin{align}
	\tilde{\mathbfit{a}}_{2\mathrm{b}} = -\frac{G M_{\mathrm{c}}}{r^2}
	\exp \left[ -A \left( \frac{r_0}{r} \right)^p \right] \mathbfit{e}_r,
	\label{eq:f2b}
\end{align}
where $r$ is the distance to the planet center, $\mathbfit{e}_r$ is the radial unit vector and
$A = 10$, $p = 8$ are parameters that ensure a quick transition between the smoothed and pure Newtonian regime
between $r \sim r_0 - 2 \, r_0$. In this work, the smoothing length $r_0$ is given by the physical radius of
the planetary core $r_{\mathrm{c}}$. It is calculated assuming a mean density of the core corresponding to a
rocky composition, $\rho_{\mathrm{c}} = 5 \, \mathrm{g}\,\mathrm{cm}^{-3}$.
We have confirmed that slightly varying the inner radius does not substantially change the outcome of the simulations.

In order to avoid artificial initial shocks, we introduce the planet's potential gently into the disc environment. We adopt the same time-dependent smoothing term as in \citetalias{Ormel2015II}. 
The injection time $t_{\mathrm{inj}}$ will vary between simulations, depending on the core mass.
After 2.5 $t_{\mathrm{inj}}$, the smoothing term has already reached 95\% of it's limit at unity.
We note that the chosen injection times are usually a few $\Omega^{-1}$ and are thus much shorter
than the actual time for the coagulation of the core.
While for massive envelopes self-gravity of the gas is
the important effect potentially leading to runaway gas accretion, we can neglect it while
the atmosphere's mass is small compared to the core mass, $M_\mathrm{env} \ll M_\mathrm{c}$.

\subsection{Numerical methods}
We solve the above described hydrodynamic equations using a modified version of the Godunov-type
hydrodynamical code \texttt{PLUTO} \citep{Mignone2007}, complemented with a module
to solve for radiation transfer in the FLD approximation \citep{Kuiper2010}.
We use a Roe solver \citep{Roe1981} to solve the Riemann problem.
The FLD equation (\ref{eq:fldeq}) is
solved using an implementation of the Generalized Minimal Residual (GMRES) method
\citep{Saad1986} including restarts for faster convergence.
For more technical details on the code, please refer to \citet{Kuiper2010}.

\section{Physical Initial Conditions}
In this section, we first describe the physical model of the protoplanetary disc that was used to
determine the environmental conditions of the planet's surroundings such as the background
mid-plane density, the local disc temperature and opacities. We then motivate the one-dimensional
simulations that we conducted for direct comparison.
\begin{table*}
    \caption{List of simulations. The first column gives the simulation name. The second column
	shows the core mass $M_\mathrm{c}$. The third column gives the injection time $t_{\mathrm{inj}}$ for the planet's gravity
	and the fourth column shows the total simulation time	$t_{\mathrm{end}}$. The spatial resolution for the domain $[r_\mathrm{c},5H] \times [0, \pi/2] \times[0,\pi]$ is given in the fifth column. Comments and references to figures are given in the sixth column.
	Standard resolution simulations were conducted
	for three different opacities.
	}
	\label{table:los}
	\centering
	\begin{tabular}{l c c c c c c c}
		\hline
		Name & $M_{\mathrm{c}} \,[M_{\oplus}]$ &
        $t_{\mathrm{inj}} \,[\Omega^{-1}]$ & $t_{\mathrm{end}} \,[\Omega^{-1}]$ & $N_r \times N_\theta \times N_\phi$ & Comments \\
			\hline
		\texttt{RT-M1-loRs} & 1 & 4 & 1000 & $~~64 \times 16 \times 32$  & resolution study, Appendix \ref{app:res}\\
		\texttt{RT-M0.1} & 0.1 & 4 & ~~500 & $~~128 \times 32 \times 64$ & lowest mass case\\
		\texttt{RT-M1} & 1 & 4 & 1000 & $~~128 \times 32 \times 64$ & Figs. \ref{fig:tautheta}, \ref{fig:McumMdot},  \ref{fig:mid_rho_vphi}, \ref{fig:Sflow-mid}, \ref{fig:MW-S-Mdot}, \ref{fig:stilder}\\
		\texttt{RT-M2} & 2 & 5 & 1000 & $~~128 \times 32 \times 64$ & Figs. \ref{fig:M2-VLO-flow}, \ref{fig:McumMdot}\\
		\texttt{RT-M5} & 5 & 5 & 1000 & $~~128 \times 32 \times 64$ & highest mass case\\
		\texttt{RT-M1-hiRs} & 1 & 4 & ~~290 & $256 \times 64 \times 128$ & resolution study, Appendix \ref{app:res}\\
		\texttt{RT-M2-hiRs} & 2 & 5 & ~~300 & $256 \times 64 \times 128$ & resolution study\\
		\hline
	\end{tabular}
\end{table*}

\begin{table*}
	\caption{List of parameters for different core masses $M_\mathrm{c}$
	in units of Earth masses. We also give the core radius, Bondi and Hill radius in Earth radii.
	The fifth column gives the dimensionless thermal mass $m = R_\mathrm{B}/H$ of the planet, which depends on orbital location
	and the local disc properties. \minc{In the last two columns, we give the radius at which we find the transition between transient and
	close to circular streamlines in the mid-plane and the envelope mass fraction $\chi_\mathrm{env} \equiv M_\mathrm{env}/M_\mathrm{c}$, both after $t=1000\, \Omega^{-1}$. Values of $\chi_\mathrm{env}$ range from results for standard opacity to \texttt{VLO} opacity.}
	}
	\label{table:pmc}
	\centering
	\begin{tabular}{c c c c c c c}
		\hline
		$M_{\mathrm{c}} \,[M_{\oplus}]$ & $r_{\mathrm{c}} \,[R_\oplus]$ &
			$R_{\mathrm{B}} \,[R_\oplus]$ & $R_{\mathrm{H}} \,[R_\oplus]$ & $m$ & \minc{$r_\mathrm{circ}/R_\mathrm{B}$}
			& $\chi_\mathrm{env}$ \\ \hline
		0.1 & $0.48$ & $1.77$ & $1.09$ & 0.04 & - & 0,018 -- 0.024\,\%\\
		1 & $1.03$ & $17.7$ & $23.4$ & 0.38 & $ 0.25$ & 1.5 -- 1.6\,\%\\
		2 & $1.30$ & $35.4$ & $29.5$ & 0.75 & $ 0.2$ & 5.2 -- 5.4\%\\
		5 & $1.77$ & $88.5$ & $40.0$ & 1.9 & $ 0.07$ & 5.1 -- 5.4\,\%\\
		\hline
	\end{tabular}
\end{table*}

\subsection{Disc model}
Consistent with a Keplerian disc, we set the initial velocity field to the state of an
unperturbed shearing flow that results from the differential rotation of the disc.
In our local approximation it is given by (see also Fig. \ref{fig:shearsheet}b)
\begin{align}
	\mathbfit{v}_{\infty}(x) = - \frac{3}{2} x\,\Omega\,\mathbfit{e}_y.
	\label{eq:v_shear}
\end{align}
In the frame where the planet is at rest, the co-rotation orbits at $x=0$ are so as well.
Gas that orbits closer to the star is overtaking the planet due to faster angular motion,
while the gas further out lags behind.

We follow \citet{Lee2014} in adopting the minimum-mass extrasolar nebula \citep[MMEN,][]{ChiangLaughlin2013}
to provide parameters for the unperturbed state of the disc.
Taken at $a=0.1 \,\mathrm{au}$, this yields values for the mid-plane gas density
${\rho_0 = 6 \cdot 10^{-6} \,\mathrm{g \, cm^{-3}}}$
and the temperature ${T_{\infty} = 1000 \, \mathrm{K}}$ \citep[Eqns. (12) \& (13)]{Lee2014}. The subscript $\infty$ stands
for unperturbed mid-plane quantities.
The obtained density is a factor of a few larger compared to the minimum-mass solar nebula
\citep[MMSN,][]{Weidenschilling1977, Hayashi1981}. The exact value of the background density should not affect the
qualitative outcome of our simulations, since we cover a range of optical depths by using different opacities. However, it might introduce an additional scaling factor for the density
profiles that we find.

Assuming an initially isothermal hydrostatic structure, the vertical density stratification
in absence of the planetary core is due to the vertical component of stellar gravity. For this
configuration the vertical density profile is given by \citep[e.g.][]{Armitage2010}
\begin{align}
	\rho_{\infty}(z) =\rho_0 \exp \left[ - \frac{1}{2} \left(\frac{z}{H}\right)^2 \right],
\end{align}
where $H$ is the vertical pressure scale-height.
Due to the isothermal nature of this configuration, the density decreases, while the entropy of the gas is increasing towards higher altitudes.

We confirmed that without including the planet's gravitational potential,
the code accurately maintains the initial equilibrium state for the full simulation time.

\subsection{Opacities}
\begin{figure}
	\includegraphics[width=.5\textwidth]{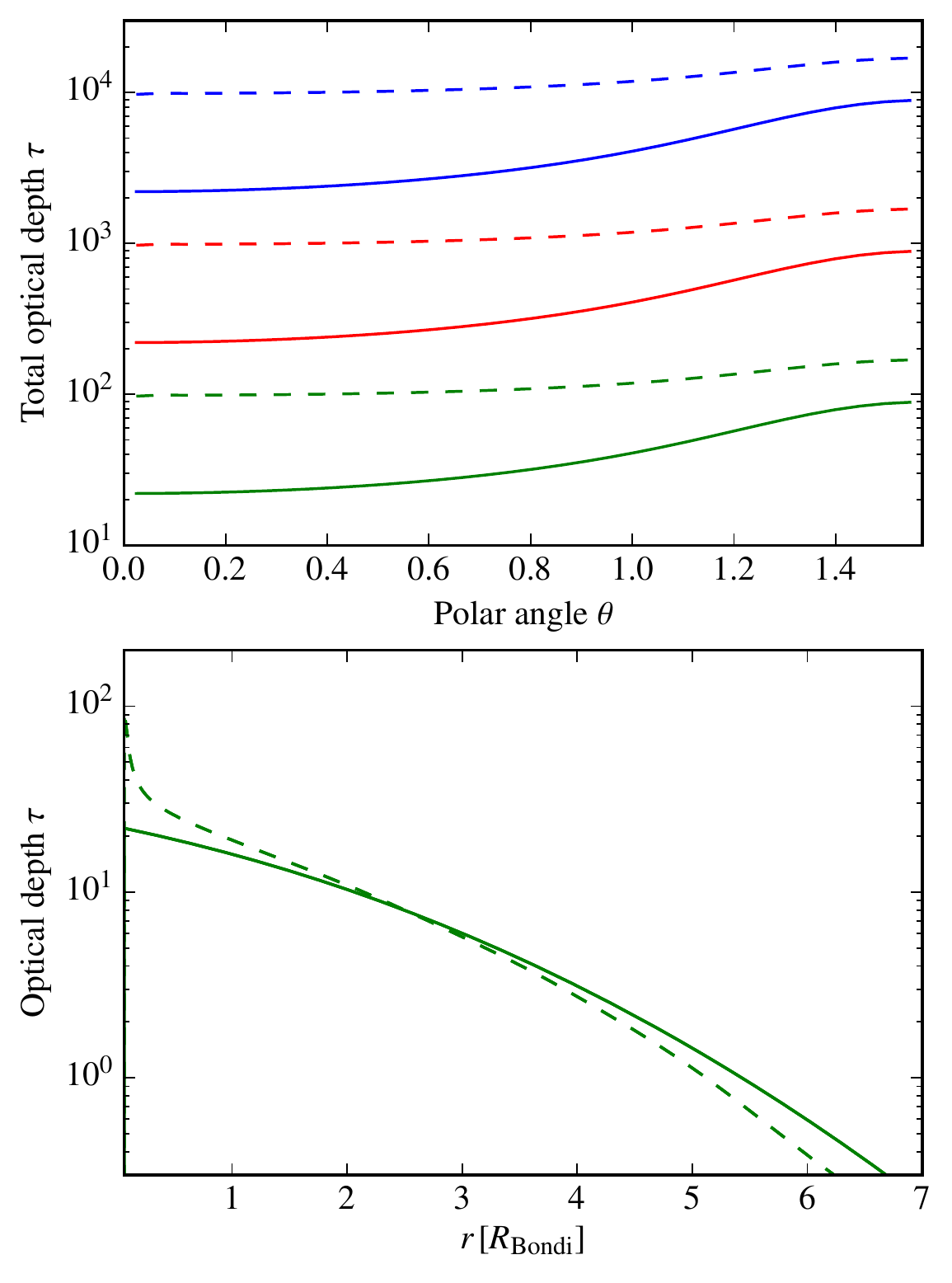}
	\caption{
	Top: Azimuthally averaged total optical depth of the simulation domain (out to $r=5\,H$) along radial rays of different polar angles $\theta$.
	Different colors correspond to the three constant values for the opacity ranging from highest (blue) to lowest (green). Solid lines
	show the state of the unperturbed disc while dashed lines correspond to the final state of simulations for a  $1\,M_\oplus$ core.
    Bottom: Optical depth as a function of radius along the polar axis ($\theta = 0$) for the very low opacity model. The meaning of line color and style is the same as in the top panel.
	}
	\label{fig:tautheta}
\end{figure}

As mentioned above, a dust-free environment is a viable scenario for super-Earth planets that we are focusing on. Due to the close-in location, the
atmospheres easily stay hot enough ($T \gtrsim 1500 \, \mathrm{K}$) to effectively evaporate dust grains, rendering
their contribution to the opacity negligible \citep[see e.g.][their Fig. 6]{LeeChiang2015}.
Another reason to consider a dust-free scenario is that the planetary core has already formed
and incorporated most of the dust in its surroundings.
Where the background disc is below the sublimation temperature of dust, coagulation of dust grains and settling
is another effective mechanism for grain removal \citep{Ormel2014, Mordasini2014}. These mechanisms result in opacities that are reduced by a few orders of magnitude
compared to typical ISM opacities.
More generally, the opacity of hot, pre-planetary atmospheres will be determined by the gas and exhibit a complex dependence on temperature, pressure and metallicity \citep{Ferguson2005,Freedman2014}. Since in this work we are mainly interested in qualitative results -- not, for example, in absolute contraction timescales -- we consider only constant Rosseland mean opacities.
Three values are explored: $\kappa_\mathrm{R} = 10^{-2}, 10^{-3}, 10^{-4}
\, \mathrm{cm}^2\, \mathrm{g}^{-1}$. Simulations with the lower opacities are tagged
as low opacity (\texttt{LO}) or very low opacity (\texttt{VLO}), respectively.
For the configuration \texttt{RT-M1} we additionally evolved a model with artificially increased opacity of $\kappa_\mathrm{R} = 10 \, \mathrm{cm}^2\, \mathrm{g}^{-1}$, which we will refer to as \texttt{VHO}.

In the top panel of Fig. \ref{fig:tautheta} we show the azimuthally averaged optical depth $\tau(\theta)$ of our
domain for radial rays as a function of the polar angle.
Solid lines mark the state of the unperturbed disc and dashed lines show the state at
the end of our simulations for an Earth-mass core. In the bottom
panel we show the optical depth along the polar direction as a function of radius.
Initially, the vertical direction ($\theta = 0$) is less optically thick
but with the contraction of the atmosphere, this feature vanishes.
Even for the lowest of our opacities of
$\kappa_{\mathrm{R}} = 10^{-4} \,\mathrm{cm}^2\,\mathrm{g}^{-1}$, our domain and the Bondi sphere are optically thick, justifying the FLD approximation.

\subsection{One-dimensional runs}
For a better understanding of the implications of the 3D nature of the flow for
the thermal evolution of the atmosphere, we also perform 1D spherically symmetric simulations using the same
code.
This allows us to look for the key differences that arise only from the
assumption of spherical symmetry, instead of comparing our findings to other 1D studies that usually
cover a different time domain and different physical effects. In these simulations we cover the same range in radius and initialize the simulation
with constant density, which is given by the mid-plane density, $\rho = \rho_0$.
We have also removed all
non-radial forces from the setup and initialize the simulation with zero velocity.
For this 1D configuration, we expect an atmosphere that is steadily cooling and contracting, because a simultaneous in- and outflow of gas is not possible.

\section{Numerical Configuration}
\subsection{Spherical Grid, Domain and Boundary conditions}
In \citetalias{Ormel2015I} we showed that already in two dimensions, a polar grid
outperforms a Cartesian grid in terms of accuracy and computational efficiency.
For the three-dimensional problem at hand, we again adopt a spherical grid
with coordinates $(r, \theta, \phi)$, centred on the planet, following \citetalias{Ormel2015II}.
Not only is the geometry more natural to the expected structure of the flow around the planetary
core, by using a linear spacing of $\log r$ in the radial dimension, it allows us to sample the
region of the atmosphere achieving high resolution while still covering a large domain in the
radial dimension. Starting at the radius of the core our local frame extends out to five pressure
scale-heights $r \in [r_{\mathrm{c}}, 5H]$. At the planet surface, we chose a reflective
boundary-condition (BC), so there is no transport of matter or radiation. For the outer
radial boundary, we set ghost-cells to the unperturbed shearing-flow state of the disc,
fixing $\rho = \rho_{\infty}(z), T = T_{\infty}$ and $\mathbfit{v} = \mathbfit{v}_{\infty}(x)$.
Parameters such as background density and temperature depend on the disc model that is
discussed above and the velocity field is given by the shearing flow in a Keplerian disc (Eqn. \ref{eq:v_shear}).

One disadvantage of this grid is the singularity at the pole $(\theta \rightarrow 0)$:
moving towards the pole, the $\mathrm{d}\phi$ length of grid cells quickly decreases,
forcing a very small time-step according to the Courant condition. In Section 2.3 of
\citetalias{Ormel2015II}, we used a linear spacing of $\cos \theta$.
This partially relaxes the constraint on the time-step
by effectively increasing the size of the cells that lie at high latitudes.
While this method is feasible for the study of low-mass planets, we realized that it is not
suited for more massive cores.
As the vertical disc stratification and vertical component of the tidal force become
relevant, the coarse polar sampling of the first grid cells quickly resulted in visible
grid artefacts during test runs. Even in test runs with doubled resolution in $\theta$,
these effects were clearly visible.
Thus, we decided to trade a longer runtime of the simulation
for higher accuracy and chose a linear sampling of the polar
angle $\theta$.
	
Since we only consider a Keplerian disc,
the equations that describe this system have a symmetry under a rotation of $\pi$ in azimuth
($\phi \rightarrow \phi + \pi$), which in Cartesian coordinates corresponds to
$(x,y,z) \rightarrow (-x,-y,z)$. In spherical coordinates it can be expressed as
${A(r,\theta,\phi) = A(r,\theta, \phi + \pi)}$ for any scalar or vector quantity $A$.
This allows us to cover the azimuthal dimension
only in the forward direction of the planet ${\phi \in [0, \pi]}$, adopting periodic
BCs. The system is also symmetric with respect to the mid-plane $(z=0)$,
so that we only cover the upper hemisphere, $\theta \in [0, \pi/2]$, that has $z > 0$ 
and take the boundary condition in the mid-plane $(\theta = \pi/2)$ to be reflective, ensuring
that no hydrodynamic or radiative fluxes cross this boundary. At the polar singularity, we use
the so-called $\pi$-boundary-condition to address ghost-cells that effectively have negative
$\theta$: $A(r,-\theta,\phi) = A(r,\theta, \phi + \pi)$ for any quantity $A$.

\subsection{Resolution and simulation time}
Compared to previous isothermal simulations, we follow the evolution of the system roughly a factor one hundred times longer than the isothermal simulations of \citetalias{Ormel2015II} in order to accurately quantify the cooling and recycling behaviour of the planet atmosphere.
For this reason, we conducted the 3D simulations in our standard
resolution, given by $N_r \times N_\theta \times N_\phi = 128 \times 32 \times 64$ cells
in the domain $[r_\mathrm{c},5H] \times [0, \pi/2] \times[0,\pi]$
for a total
simulation time of $t_{\mathrm{end}} = 1000 \,\Omega^{-1} \simeq 5 \, \mathrm{yr}$.
One exception to the rule is simulation \texttt{RT-M1-VLO}, which we
evolved up to $t=2000 \, \Omega^{-1}$ in order to judge the convergence to a steady state.
As an example for an Earth-mass core, our standard resolution allows for 67 cells inside the
Bondi radius in the radial direction.
In a resolution study (see Appendix \ref{app:res}), we conducted simulations with half/twice as many cells (\texttt{loRs}/\texttt{hiRs})
in each dimension in order to find and understand resolution effects.
However, \texttt{hiRs} runs were computationally too costly to cover the same time domain.
A list of simulations can be found in Table \ref{table:los}.

\section{Results}
In this section we present and analyse the results of the radiative hydrodynamical models.
First we discuss the general structure of the atmospheric region and dynamical interaction with flows from the disc in
Section \ref{subs:flowstructure}.
We then describe the thermal properties, focusing on the transport of entropy by advection and radiation in Section \ref{subs:entropytransport}.
A direct comparison between 1D and 3D simulations is made in Section \ref{subs:1d3d}.

\subsection{Three-dimensional flow structure}
\label{subs:flowstructure}
\begin{figure*}
	\centering
	\includegraphics[width=17cm]{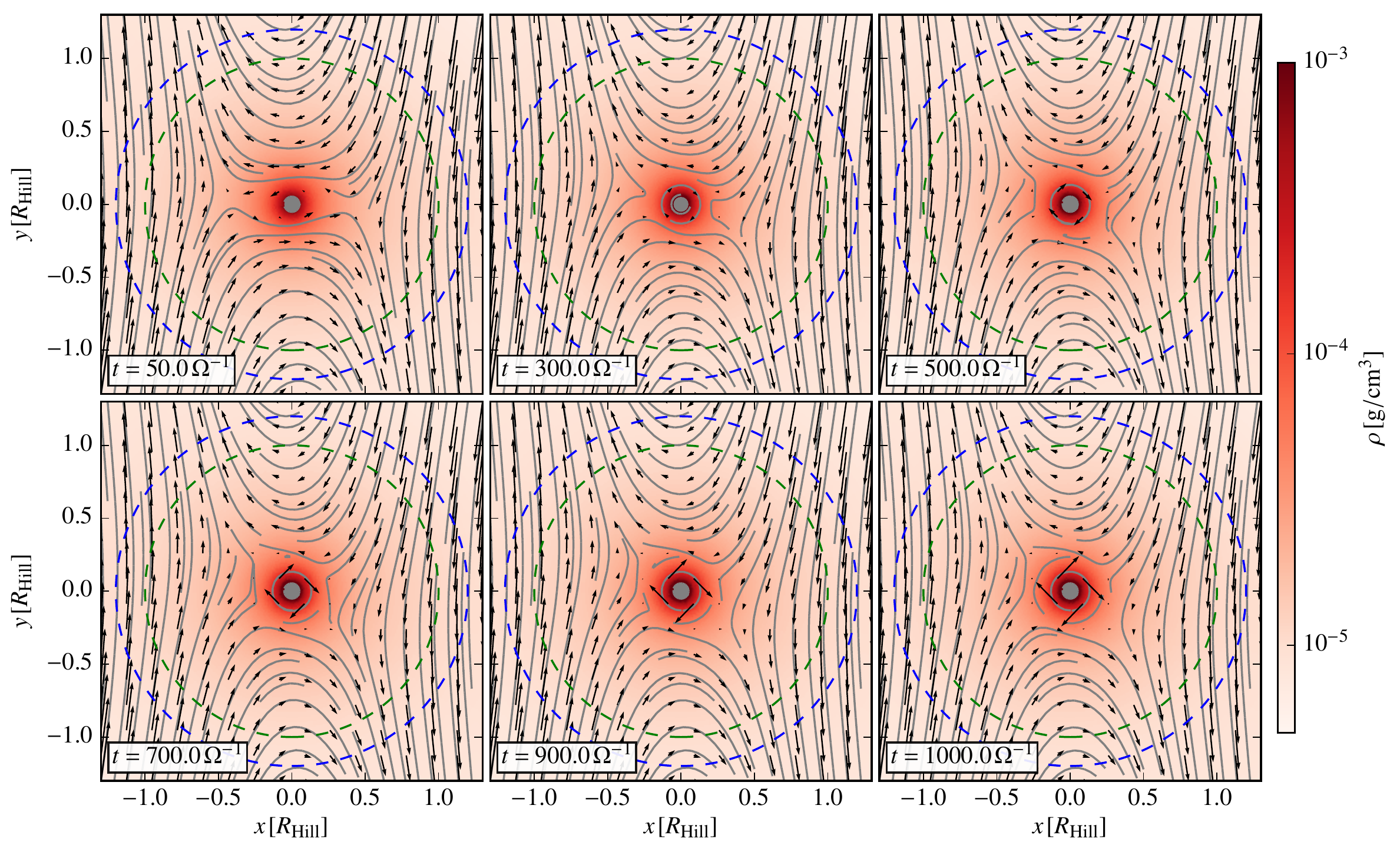}
	\caption{Time evolution of the density and the flow structure in the mid-plane for simulation \texttt{RT-M2-VLO}.
	Dashed circles mark the Bondi (blue) and Hill (green) radius.
	The arrows show gas velocity and the grey lines are corresponding streamlines. By symmetry the vertical velocity component
    vanishes here.}
	\label{fig:M2-VLO-flow}
\end{figure*}

\begin{figure*}
	\centering
	\includegraphics[width=17cm]{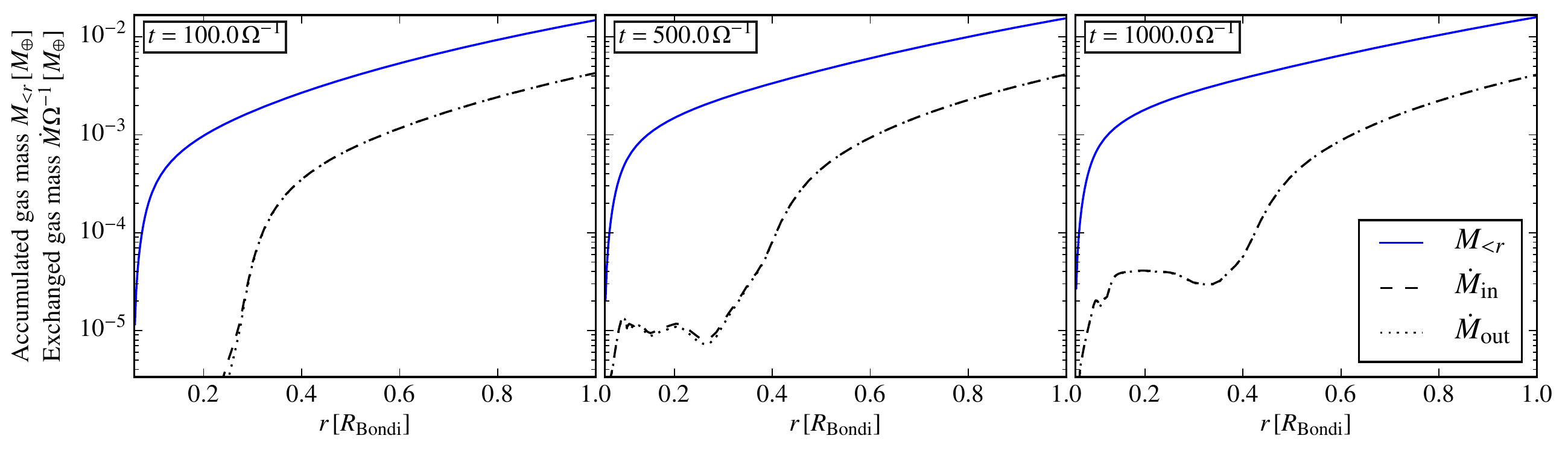}\\
	\vspace{-.88cm}\hspace{0.01cm}
	\includegraphics[width=17cm]{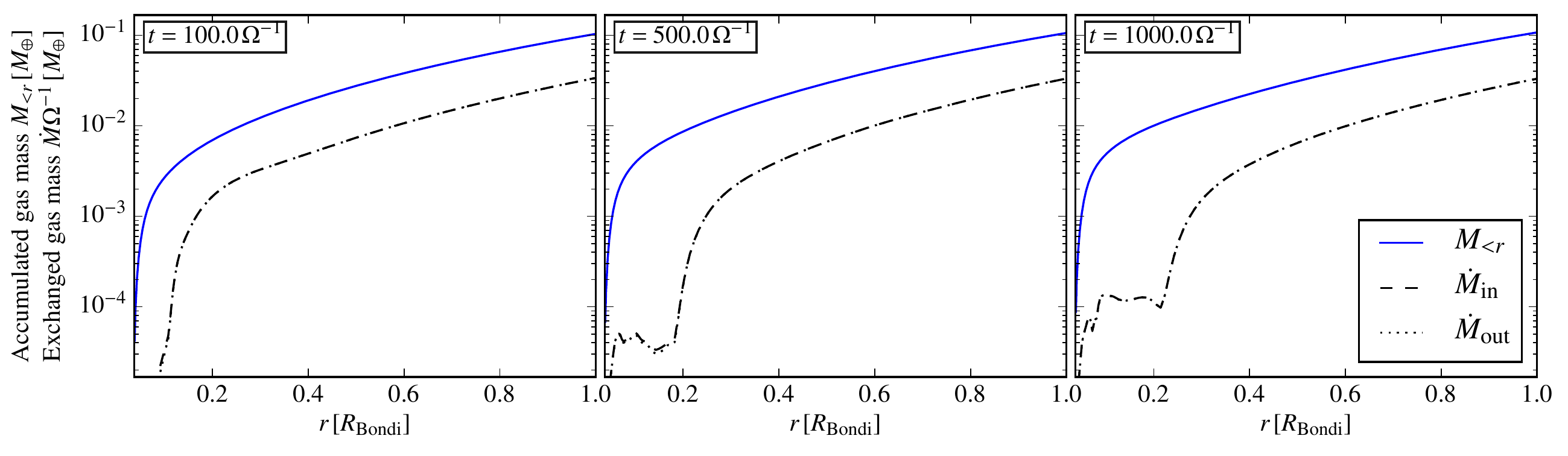}
    \caption{Time series of cumulative gas mass $M_{<r}$ (solid blue) inside radius $r$ together with accretion rates corresponding to in- (dashed) and outflow (dotted)
	scaled with the Keplerian frequency.
	Top: \texttt{RT-M1-VLO}, bottom: \texttt{RT-M2-VLO}. While mass fluxes are high in the
    outer regions of the Bondi sphere, they decreased closer in.
	Incoming and outgoing streams compensate each other, resulting in much smaller net accretion.
    Over time the inner regions allow for more exchange of gas.
    The pileup of gas close to the planet due to cooling
    and contraction is visible as an increase in $M_{<r}$ close in.
	}
	\label{fig:McumMdot}
\end{figure*}

\begin{figure}
	\centering
	\includegraphics[width=.5\textwidth]{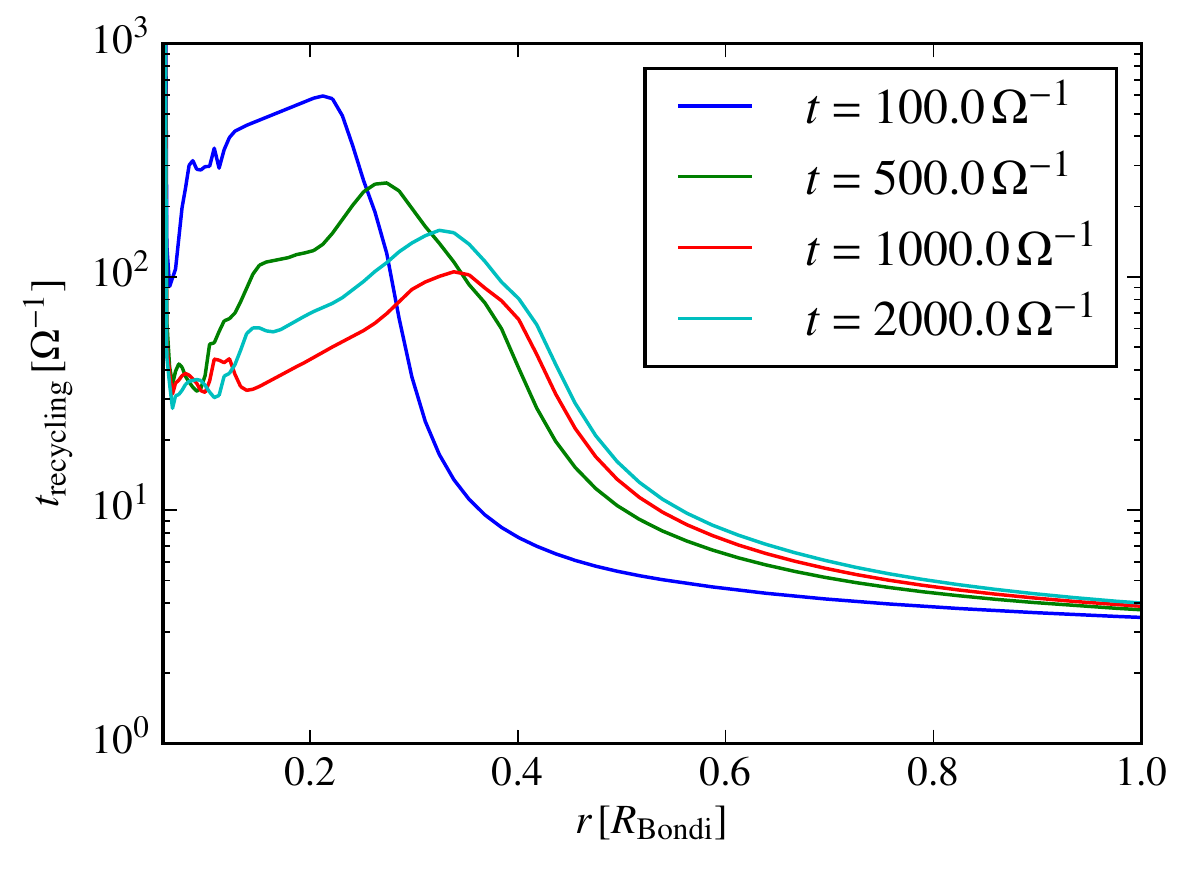}\\
    \caption{Time evolution of the recycling time $t_\mathrm{recycling}$ (Eqn. \ref{eq:trecycling}) for \texttt{RT-M1-VLO} as a function of radius. While gas in the outer regions is quickly replenished, the time-scale on which recycling occurs closer in is usually larger, as the atmosphere becomes denser and mass fluxes reduce (compare Fig. \ref{fig:McumMdot}).
	}
	\label{fig:t_recycle}
\end{figure}

\begin{figure}
	\centering
	\includegraphics[width=\columnwidth]{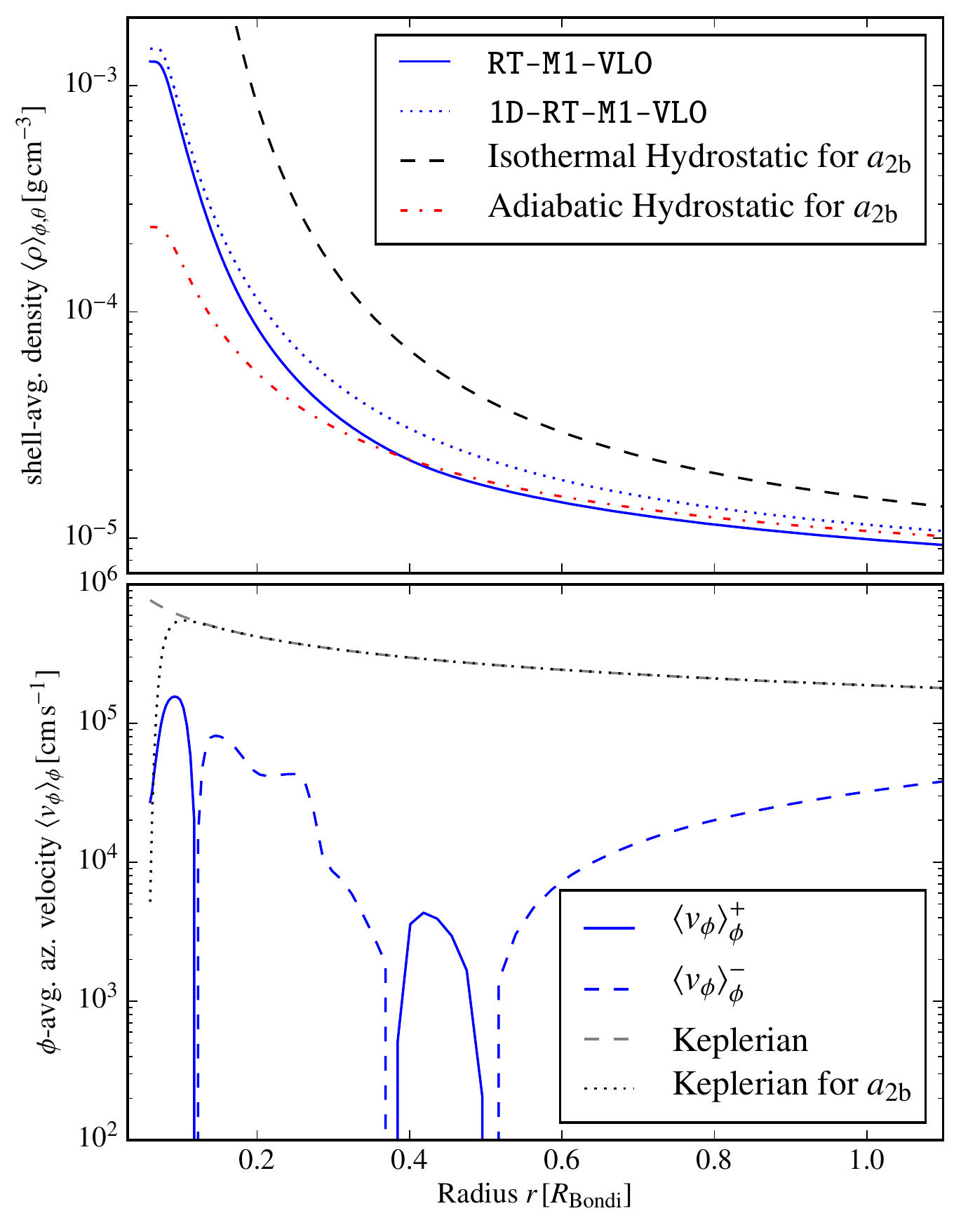}
	\caption{Simulation results for \texttt{RT-M1-VLO} at $t = 2000\,\Omega^{-1}$. Top: shell-averaged density
	profile (blue) for 3D (solid) and 1D (dotted) runs compared to analytical hydrostatic solutions employing an isothermal (black dashed) and adiabatic (red dash-dotted) EoS.
	The simulation results closely follows the adiabatic profile, deviating only close in.
	Bottom: azimuthally averaged azimuthal velocity  in the mid-plane (blue) split into positive (solid) and negative (dashed) component.
	We also show rotational profiles of Keplerian rotation in a Newtonian (grey dashed) and in our smoothed
	potential (grey dotted). We recognize that the retrograde rotation from Fig. \ref{fig:Sflow-mid} is well
	sub-Keplerian.
	}
	\label{fig:mid_rho_vphi}
\end{figure}

For all our simulations we find that the ``atmosphere'' -- material within the Bondi and Hill sphere -- is connected to the disc by flows both entering and leaving: recycling always operates.
In Fig. \ref{fig:M2-VLO-flow} we show the time evolution
of the flow and density structure in a mid-plane section for a two Earth mass core as a typical example.
The blue and green dashed circles mark the Bondi and Hill radius, respectively.
At $\vert x \vert \sim R_\mathrm{B}$ we recognize the shearing flow, streamlines further in are bent and pass closer to
the core. We also recognize typical horseshoe streamlines in the co-rotation region around $x=0$ which are slightly skewed
and not symmetric with respect to the $y$-axis.

The resulting mass fluxes $\dot{M}_{\mathrm{in/out}}(r)$ through a spherical surface
at radius $r$ and cumulative mass $M_{<r}$ inside this sphere
are plotted in Fig. \ref{fig:McumMdot} for two simulations.
\footnote{Different than in \citetalias{Ormel2015II}, we directly saved the mass fluxes across cell interfaces as simulation data, instead
of calculating them from cell-centred velocities and density in post-processing.}
The top panel shows results for
\texttt{RT-M1-VLO}, the bottom panel for \texttt{RT-M2-VLO}. Mass fluxes are higher in the outer
parts of the Bondi sphere and drop off towards the center. The rates of in- and outflow
match very closely, resulting in a net accretion that is orders of magnitudes smaller.
By taking the ratio of mass fluxes and the cumulative mass, we can define a recycling
time-scale
\begin{equation}
	t_{\mathrm{recycling}} \equiv \frac{M_{<r}}{\dot{M}_{\mathrm{in}}},
	\label{eq:trecycling}
\end{equation}
similar to that in \citetalias{Ormel2015II}. However, here we do not sort the mass fluxes by
the minimum radius they reach by integrating their streamlines, but look at the local mass flux. In other words, this does not tell us directly, how deep the
streams penetrate the atmosphere but still gives us an understanding of how large the
mass exchange is at each radius. This quantity is shown in Fig. \ref{fig:t_recycle} for \texttt{RT-M1-VLO} at different times as a
function of radius. For $r \gtrsim 0.4 R_\mathrm{B}$, we have
$t_{\mathrm{replenish}} \sim 1 - 10 \,\Omega^{-1}$ i.e., days, while in the inner regions we usually find
$t_{\mathrm{recycling}} \sim 100 \,\Omega^{-1}$ i.e., years. With time, the recycling time in the outer part slightly increases, while we see a decrease close in. The latter trend has to be taken with care however, since due to our local definition of the recycling time, circulation of material that has some radial component will affect this quantity, without effectively transporting in gas from the disc. Thus, recycling times might be underestimated for layers close to the core.
The fraction of a Bondi radius at which this transition in recycling time happens is smaller for a two Earth mass core compared
to a one Earth mass.
This means that the layers of the envelope are not isolated but steadily exchange gas on a larger scale than the net accretion that is allowed for by radiative cooling.

Over time, due to the atmosphere's contraction, mass accumulates close to the core.
As angular momentum is transported to its inner parts, rotation increases. The retrograde nature of this rotation is different to the results of \citetalias{Ormel2015II}, where we found outward spiralling streams with a prograde sense in the mid-plane.

To assess the importance of rotation we show the shell averaged density profile and azimuthally averaged azimuthal velocity in Fig. \ref{fig:mid_rho_vphi}. The top panel compares the simulation results (blue) of
3D (solid) and 1D (dashed) models with analytical hydrostatic solutions
corresponding to the planet's smoothed gravity, adopting an isothermal (black dashed) and adiabatic (red dash-dotted) equation of state.
The gravitational smoothing which causes the vanishing density gradient at the inner boundary, hardly affects the atmosphere structure.
Overall, the simulation results agree with an adiabatic structure with deviations close in, where the atmosphere has contracted the most.
We confirm that right after the injection of the planet, before the atmosphere has a chance to cool, results of \texttt{RT} runs agree closely with adiabatic profiles. For higher opacity runs, this remains true for longer times than \texttt{VLO} runs. Our very high opacity test \texttt{VHO} maintains virtually constant adiabatic structure up to
$t=1000\,\Omega^{-1}$.

The bottom panel shows the azimuthal
velocity (blue) with positive/negative values drawn as a solid/dashed line. Comparing the simulation results to velocities of Keplerian rotation,
we find that the any of the rotation seen in Fig. \ref{fig:Sflow-mid} is sub-Keplerian and dynamically unimportant in supporting the
atmosphere. Clearly, there is no evidence for a circumplanetary disc that is often found in simulations
of very massive planets \citep[e.g.][]{AyliffeBate2012,TanigawaEtal2012,SzulagyiEtal2016}, at this point. The sign change in azimuthal velocity from positive (prograde) close to the planet
surface to negative (retrograde) at the point, where gravitational smoothing kicks in is most
likely caused by this simplified treatment of gravity, since varying the smoothing parameter
$r_\mathrm{smooth}$ slightly changed its location (see Appendix \ref{app:res}).

In Table \ref{table:pmc}, we give the radius $r_\mathrm{circ}$, where we find the transition between transient and circular, (quasi)bound
streamlines in the mid-plane as a fraction of the Bondi radius. It is defined as the largest radius inside the atmosphere, where the azimuthally averaged ratio of radial
to azimuthal velocities falls below a threshold of ten percent, so where $\langle \vert v_r/v_\phi \vert \rangle_\phi \leq 0.1$.
Although the choice of this threshold value is somewhat arbitrary, it helps in identifying a trend for the dependence of this radius on
the core mass. We emphasize that this crude analysis is limited to the mid-plane and does not directly tell us whether there is still mass-exchange happening at higher latitudes, which we will address in Section \ref{subs:entropytransport}.
We also give the envelope mass fraction
$\chi_\mathrm{env} \equiv M_\mathrm{env}/M_\mathrm{c}$ at $t=t_\mathrm{end}$ \minc{(i.e. the mass inside $R_\mathrm{B,H}$)} for different core masses \minc{for standard and
\texttt{VLO} opacity}. While the circulation radius decreases with increasing core mass, the envelope mass fraction increases and is typically
of several percent for super-Earth mass cores. \minc{The envelope masses show little dependence on the adopted opacity at this point. Since we consider a gas-rich environment, even fully adiabatic atmospheres have significant mass fractions of some percent.}

We note however, that generally our simulations have not yet reached steady state.
Thus, by comparing envelopes for different core masses at a fixed time we are viewing them at different stages of their thermal evolution.

\subsection{Entropy transport}
\label{subs:entropytransport}
\begin{figure*}
	\centering
	\includegraphics[width=17cm]{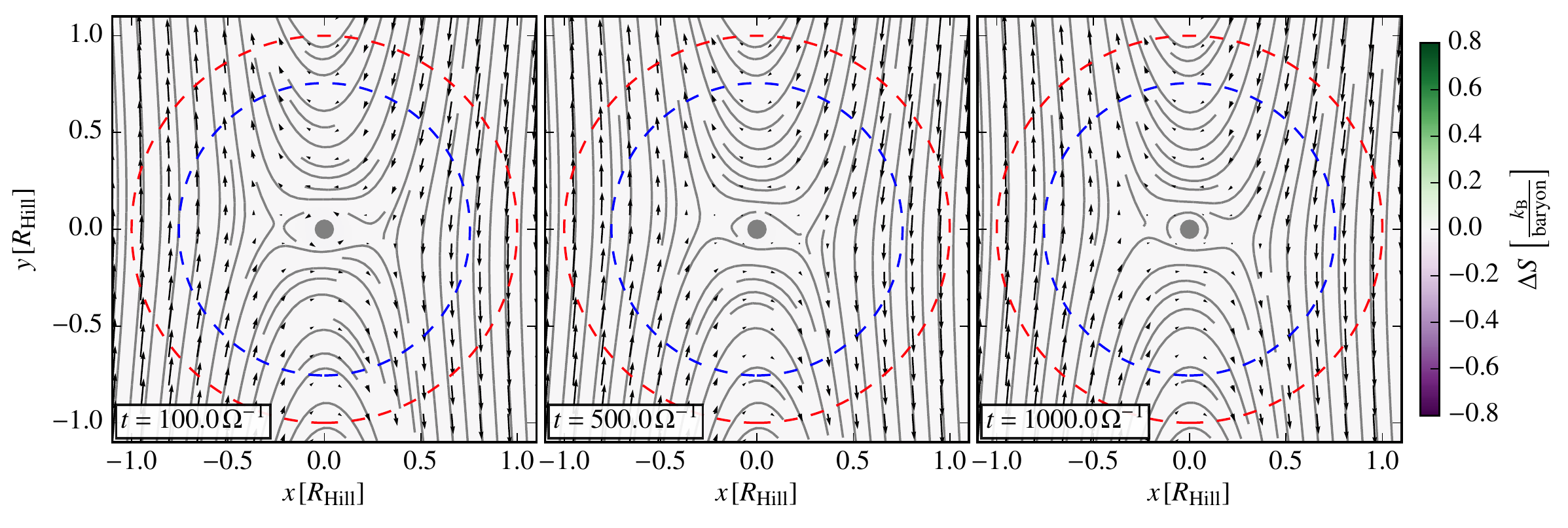}\\
	\vspace{-0.5cm}
	\noindent\makebox[\linewidth]{\rule{\textwidth}{0.4pt}}
	\centering
	\includegraphics[width=17cm]{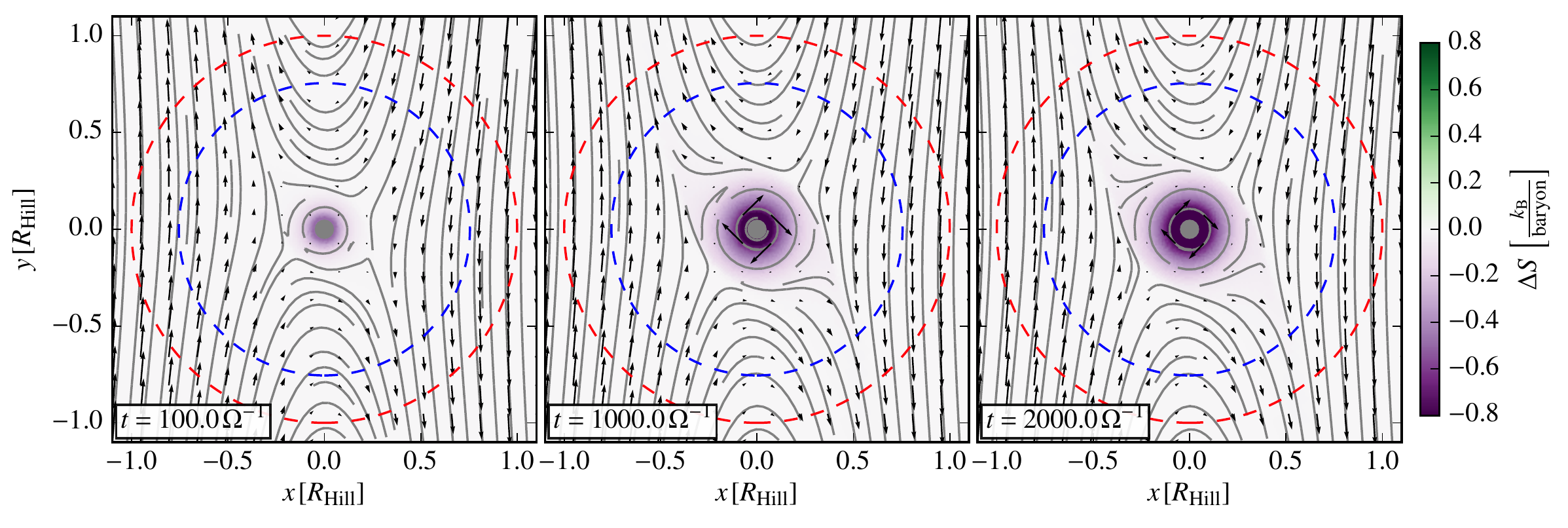}\\
	\vspace{-0.6cm}
	\includegraphics[width=17cm]{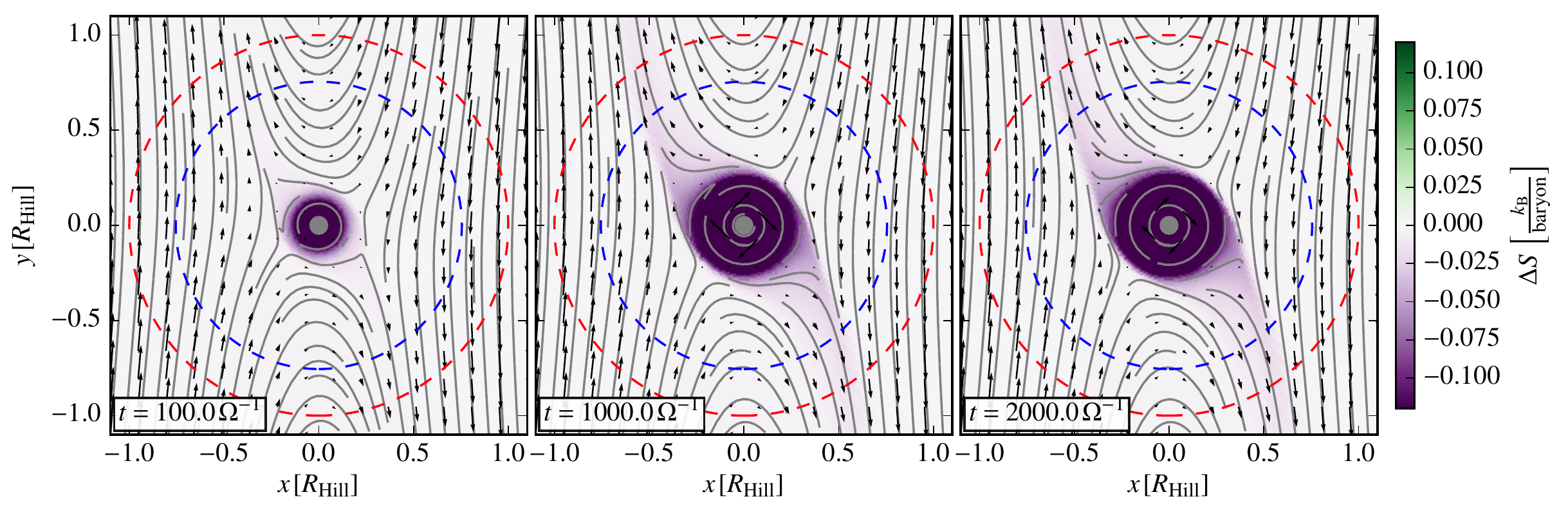}
	\caption{Mid-plane flow structure for \texttt{RT-M1-VHO} (top) and \texttt{RT-M1-VLO} (bottom), similar to Fig.
	\ref{fig:M2-VLO-flow}. Dashed circles mark the Bondi (blue) and Hill (red) radius.
	Coded with colors is diffrence in specific entropy to the unperturbed
	mid-plane value. The two bottom rows show the same data, but with different scaling
	of the colormap. The bottom row emphasizes the mixing of gas and resulting lower entropy outflow
	in the mid-plane.
	}
	\label{fig:Sflow-mid}
\end{figure*}
\begin{figure*}
	\centering
	\adjincludegraphics[trim={0 {.43\height} 0 0}, clip, width=14cm]
		{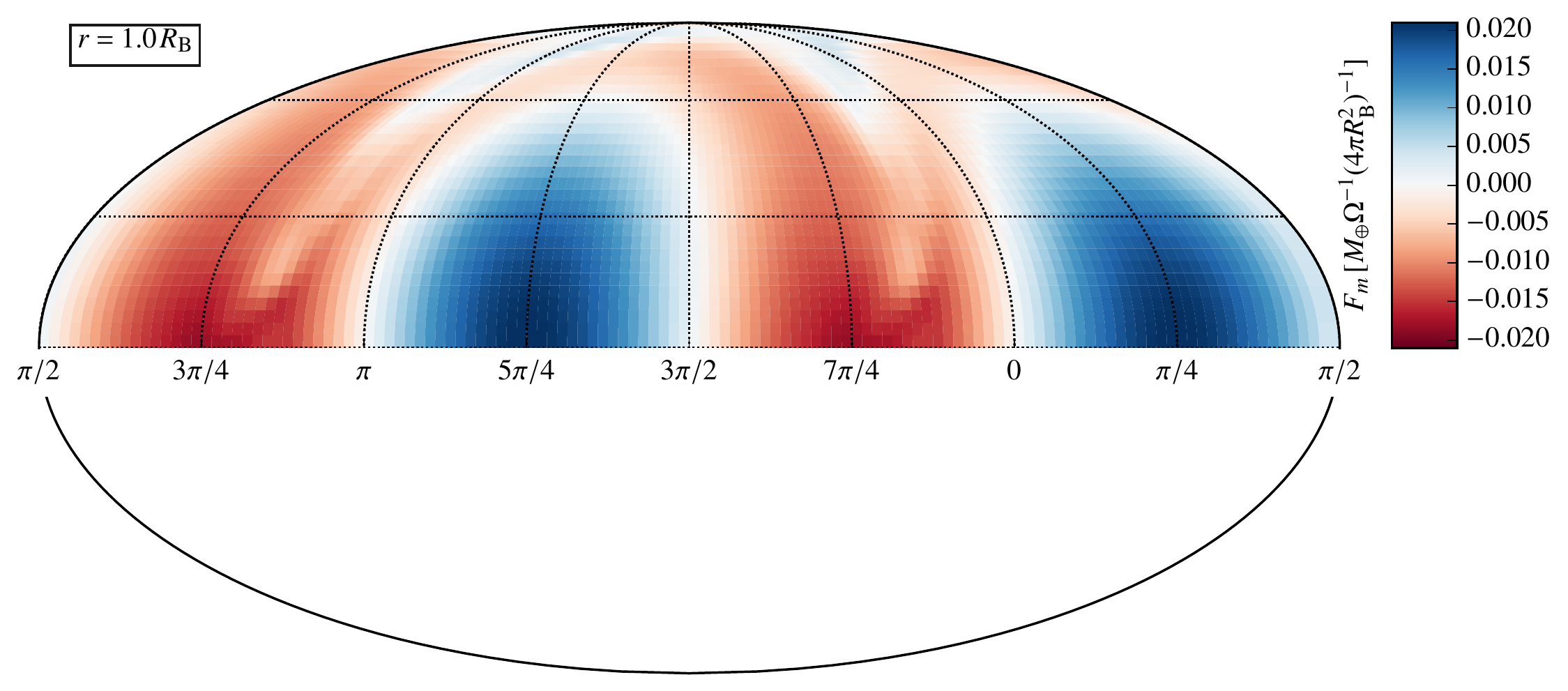}\\
	\adjincludegraphics[trim={0 {.43\height} 0 0}, clip, width=14cm]
		{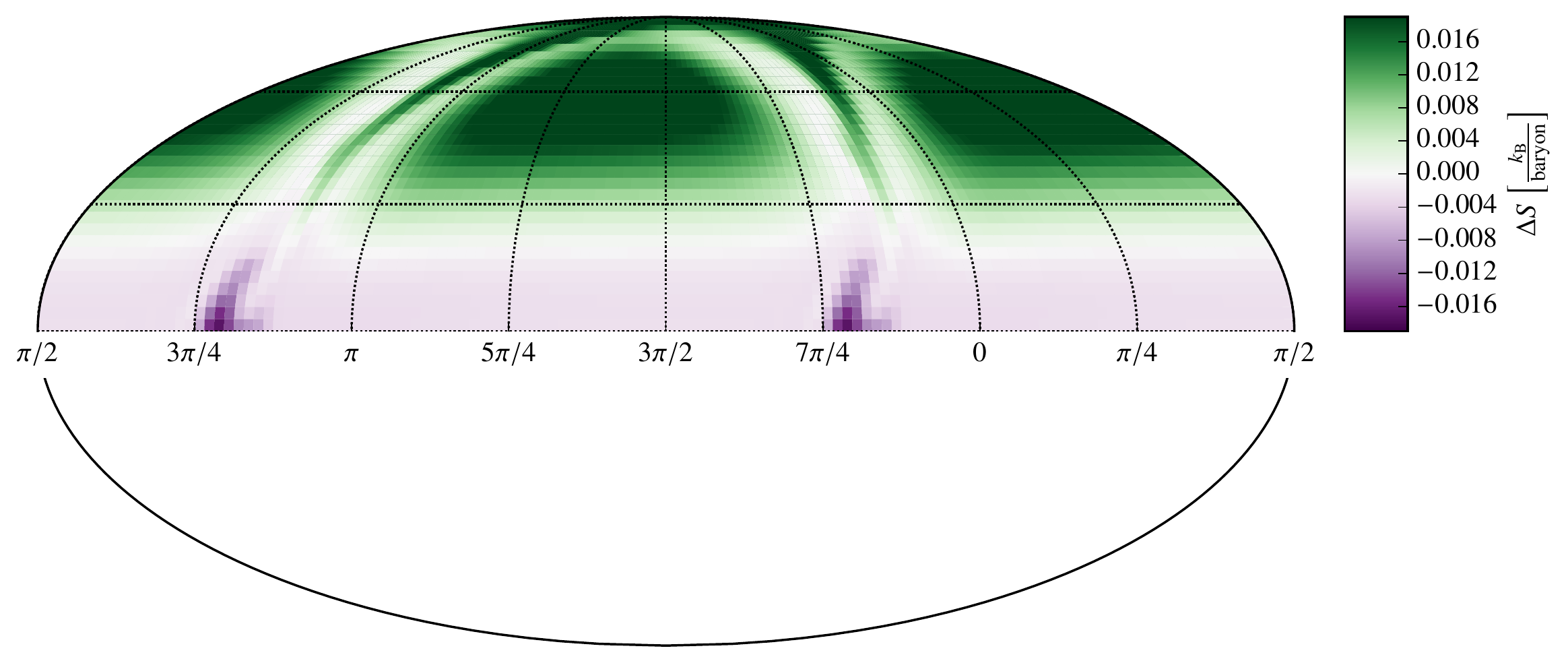}\\
		\noindent\makebox[\linewidth]{\rule{\textwidth}{0.4pt}}
	\adjincludegraphics[trim={0 {.43\height} 0 0}, clip, width=14cm]
		{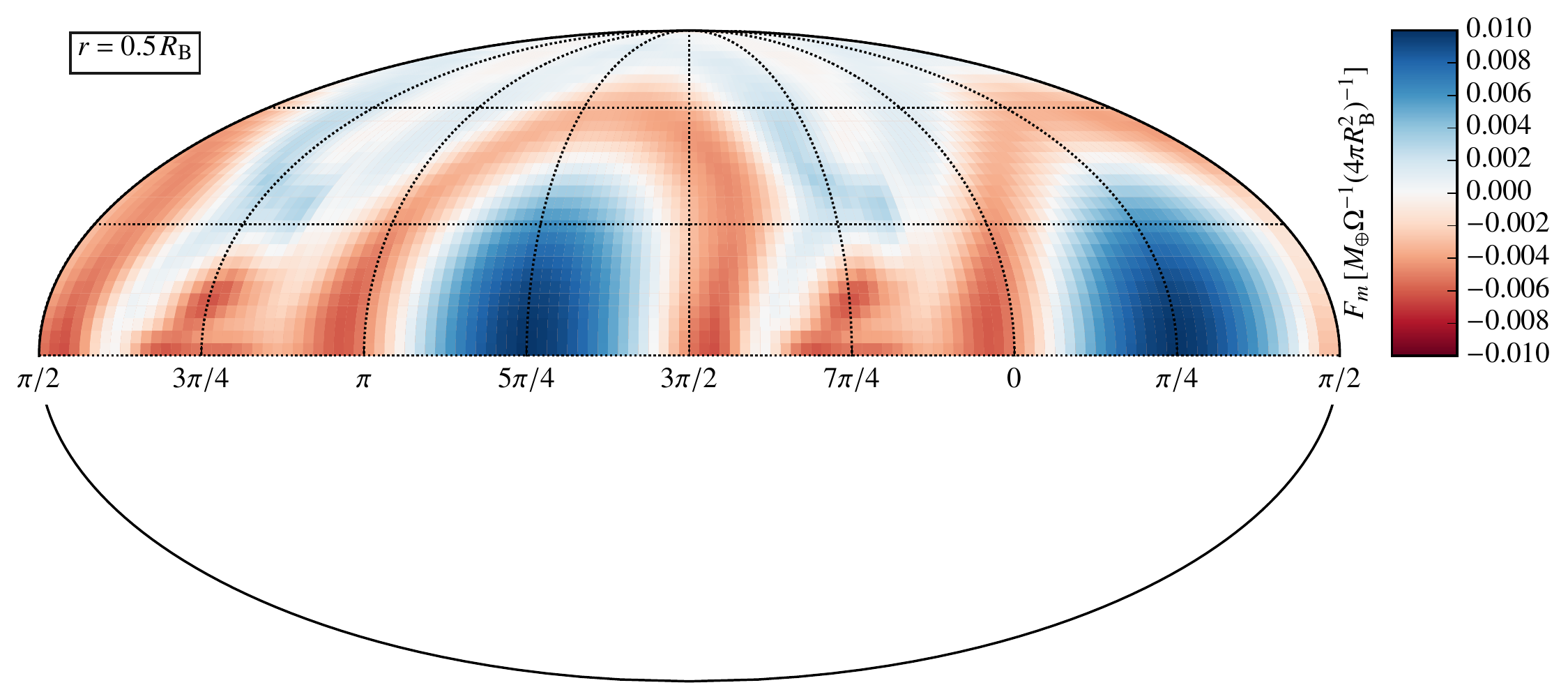}\\
	\adjincludegraphics[trim={0 {.43\height} 0 0}, clip, width=14cm]
		{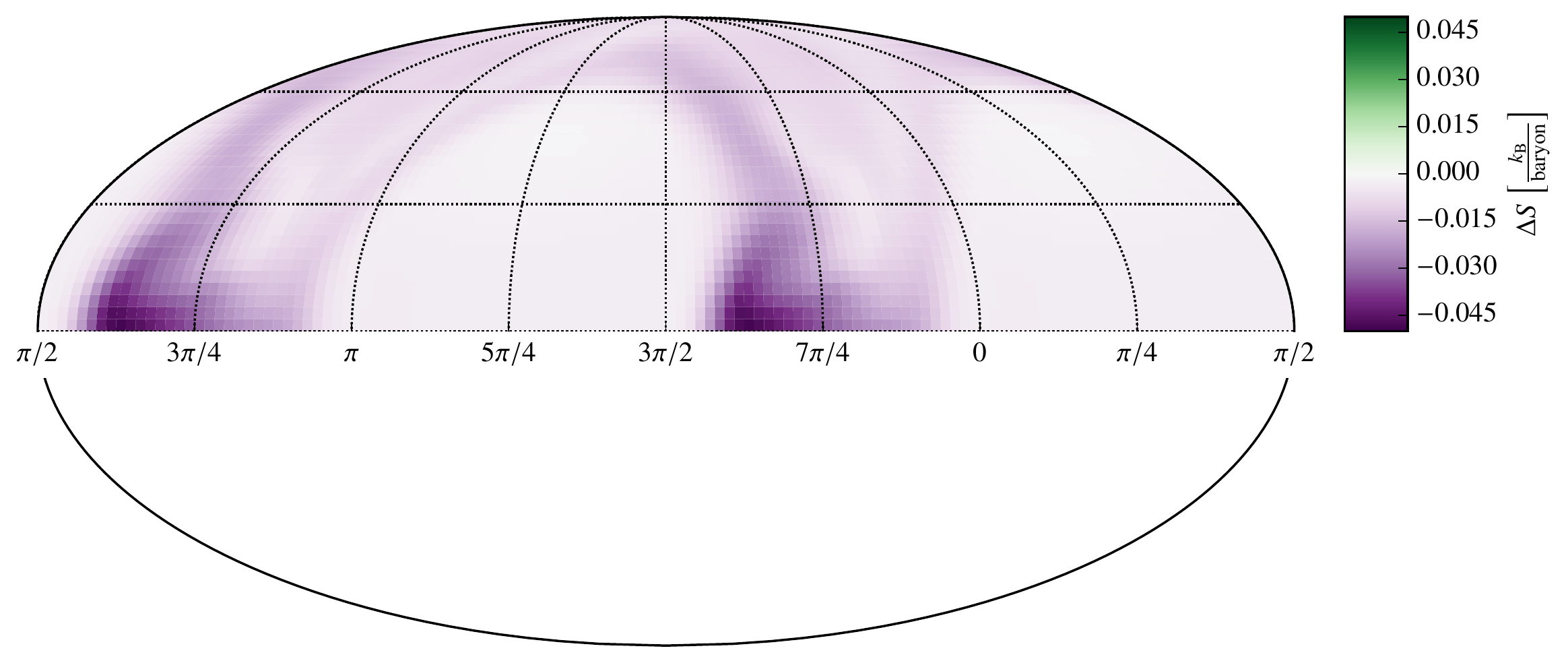}\\
		\noindent\makebox[\linewidth]{\rule{\textwidth}{0.4pt}}
	\adjincludegraphics[trim={0 {.43\height} 0 0}, clip, width=14cm]
		{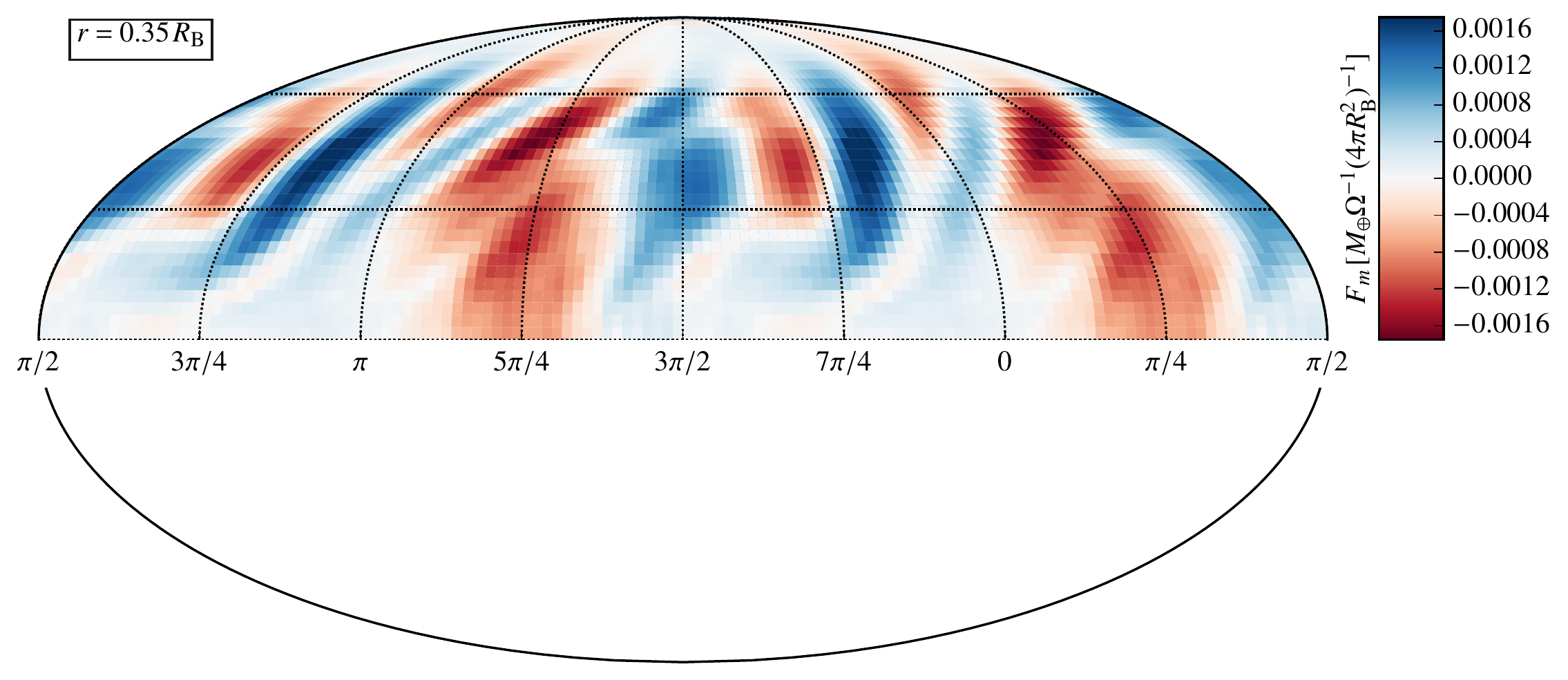}\\
	\adjincludegraphics[trim={0 {.43\height} 0 0}, clip, width=14cm]
		{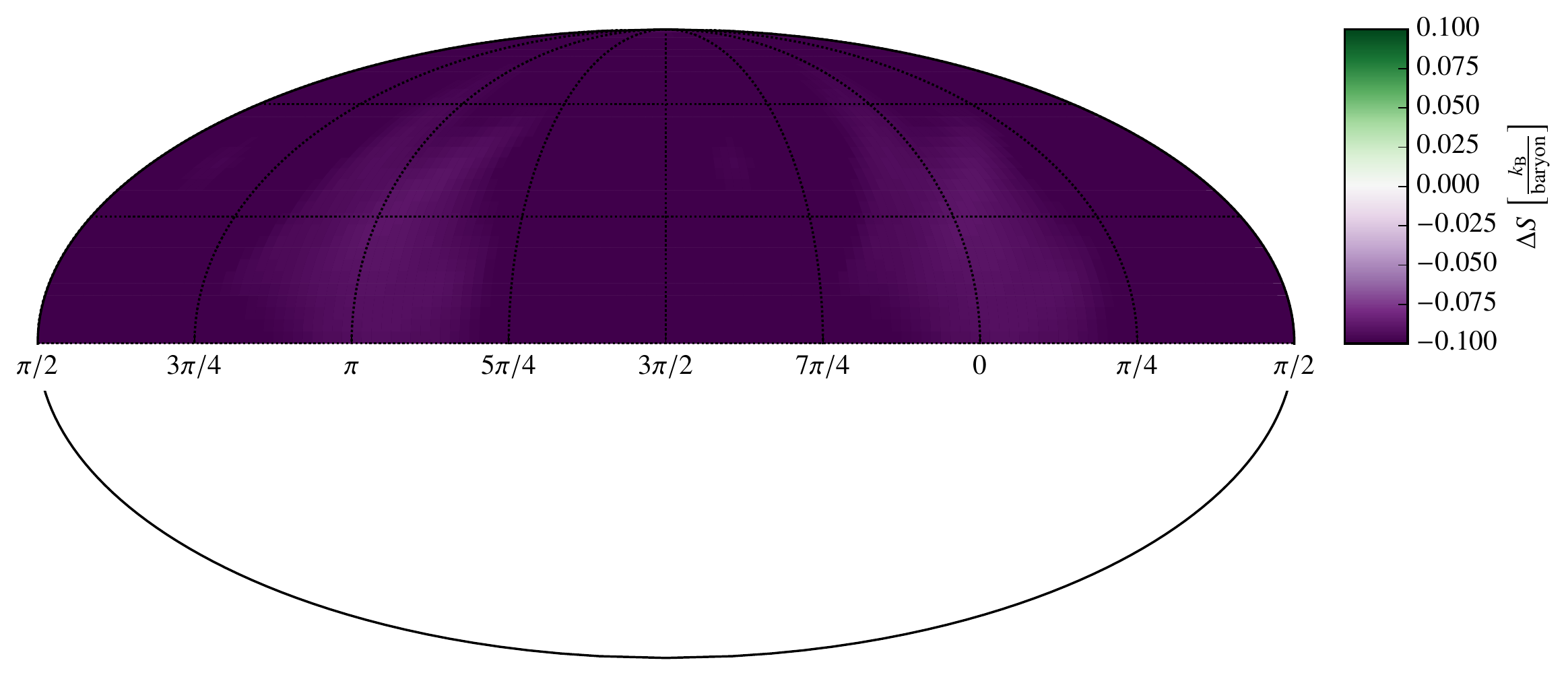}
    \caption{
	Mass flux $F_m$ and specific entropy $\Delta S$ on spherical surfaces at $r = R_\mathrm{B}$ (top),
	$r = 0.5 \, R_\mathrm{B}$ (middle) and $r = 0.35 \, R_\mathrm{B}$
	(bottom) for \texttt{RT-M1-VLO} at $t = 1000\,\Omega^{-1}$.
	All quantities are averaged over $1\,\Omega^{-1}$.
	The mass flux has similar units as accretion rates above, but is additionally scaled with the surface
	of the Bondi sphere. It is defined such that positive values (blue) correspond to inward
	mass flux.
	We only show the upper hemisphere because of the mid-plane symmetry. The longitude of $\phi = 3 \pi / 2$ corresponds to
	the view that is following the planets orbital direction.
	On the surface of the Bondi sphere, we still recognize the vertical stratification in entropy
	giving $\Delta S > 0$ at high altitudes. There, we also find a clear trend for leaving
    gas (red) having lower entropy compared to gas entering the atmosphere (blue).
    }
	\label{fig:MW-S-Mdot}
\end{figure*}

The implications
of the previously-discussed recycling on the cooling of the
planet's atmosphere can best be described in terms of entropy.
In the following we
express the difference in specific entropy in units of $k_\mathrm{B}$ per baryon as
\begin{align}
	\frac{\Delta S}{k_\mathrm{B}/\mathrm{baryon}}  = \frac{1}{(\gamma - 1) \mu} \ln \frac{P/P_\infty}{(\rho/\rho_\infty)^\gamma},
\end{align}
where $k_\mathrm{B}$ is Boltzmann's constant and
again the subscribed quantities are those of the unperturbed mid-plane.
In Fig. \ref{fig:Sflow-mid} we show the mid-plane flow for \texttt{RT-M1-VHO} and \texttt{RT-M1-VLO}, color coded by $\Delta S$.
The two bottom rows show the same data with different ranges of the colormap. A clear distinction between two regions can be made: (i) an outer near-isentropic region where $S$ is the same as that of the disc; and (ii) an inner core where $S$ is clearly lower.
In the outer regions, because of continuous exchange of gas from the disc, most of the gas is able to maintain entropy close to the initial value.
This picture remains almost unchanged between the snapshots at $t = 500 \,\Omega^{-1}$ and $t = 1000 \,\Omega^{-1}$, suggesting rapid convergence to steady state. However, the evolution proceeds differently within the hot, close-in part of the atmosphere, where gas cools efficiently by radiation. Initially, the extent of the dark area of low entropy gas grows.
With time however, the inner region is also affected by shearing and horseshoe orbits, which inject high entropy gas while removing low entropy gas from the very inner regions. This process is emphasized in the bottom panel of Fig \ref{fig:Sflow-mid}. Comparing the results for different opacities, we see that more efficient cooling (and thus lower entropy) promotes the transition towards more circular -- and arguably more bound -- motion around the core.
We find a similar structure for simulation \texttt{RT-M2-VLO}.

We emphasize however that 3D effects are key. Fig. \ref{fig:MW-S-Mdot} displays the mass flux $F_m$ (in units of Earth masses per inverse
orbital frequency per surface of the Bondi sphere) and the specific entropy
$\Delta S$ on a spherical surface at $r = 1 \, R_\mathrm{B},\, 0.5 \, R_\mathrm{B}$ and $0.35 \, R_\mathrm{B}$. All quantities are averaged over a time of one $\Omega^{-1}$. Due to the symmetry of our model,
we describe the longitude $\phi$ of only one feature, the other will be found at $\phi' = \phi + \pi$.
The longitude of $\phi = 3\pi /2$ corresponds to looking at the planet from the $-y$ direction,
following its orbital direction.
On the surface of the Bondi sphere the mass flux through the atmosphere is dominant in the
mid-plane, where mainly the shearing and horseshoe flow is passing perpendicularly through it.
The intensity of
the mass flux is highest there and reduces towards higher latitudes, where the outward
flow dominates. Comparing the flow pattern to the entropy plot below, we recognize the
low-$S$ outflow from Fig. \ref{fig:MW-S-Mdot} in the mid-plane.
For this feature at $\phi \simeq 3\pi /4$, the lowest entropy flows are close to the mid-plane.
However, it extends vertically around this longitude, so that even at high latitudes the
out-flowing streams show lower entropy compared to their unperturbed state.
In general we find that on the Bondi sphere, regions of accretion overlap with regions of higher
entropy compared to areas of outflow.
Making the same surface cut at $r = 0.5 \, R_\mathrm{B}$, we find a similar structure, with the
main accretion streams having $\Delta S \sim 0$, the disc's mid-plane entropy. At this distance from the core mass fluxes are still the
highest close to the mid-plane, while at higher latitudes they virtually vanish.
Closer in at $r = 0.35 \, R_\mathrm{B}$, the mass flux
in the mid-plane almost vanishes, corresponding to the circular streamlines that we find there,
allowing only for very little outflow. However, the inner region is still exchanging gas at
higher latitudes.
Gas is leaking the strongest at $\phi = \pi /4$, which
corresponds to the line of $x = y$, regions of highest accretion are shifted by $\pi/2$.
In contrast to the outer regions, at $r=0.35 \, R_\mathrm{B}$, we don't find any gas that has $\Delta S \simeq 0$ and
the correlation between the direction of the flow and its entropy
has blurred.
Together with the reduced mass flux (Fig. \ref{fig:McumMdot}), this suggests that advective
entropy transport is less efficient by this radius.

\subsection{Comparison 3D vs. 1D}
\label{subs:1d3d}
\begin{figure*}
	\centering
	\includegraphics[width=17cm]{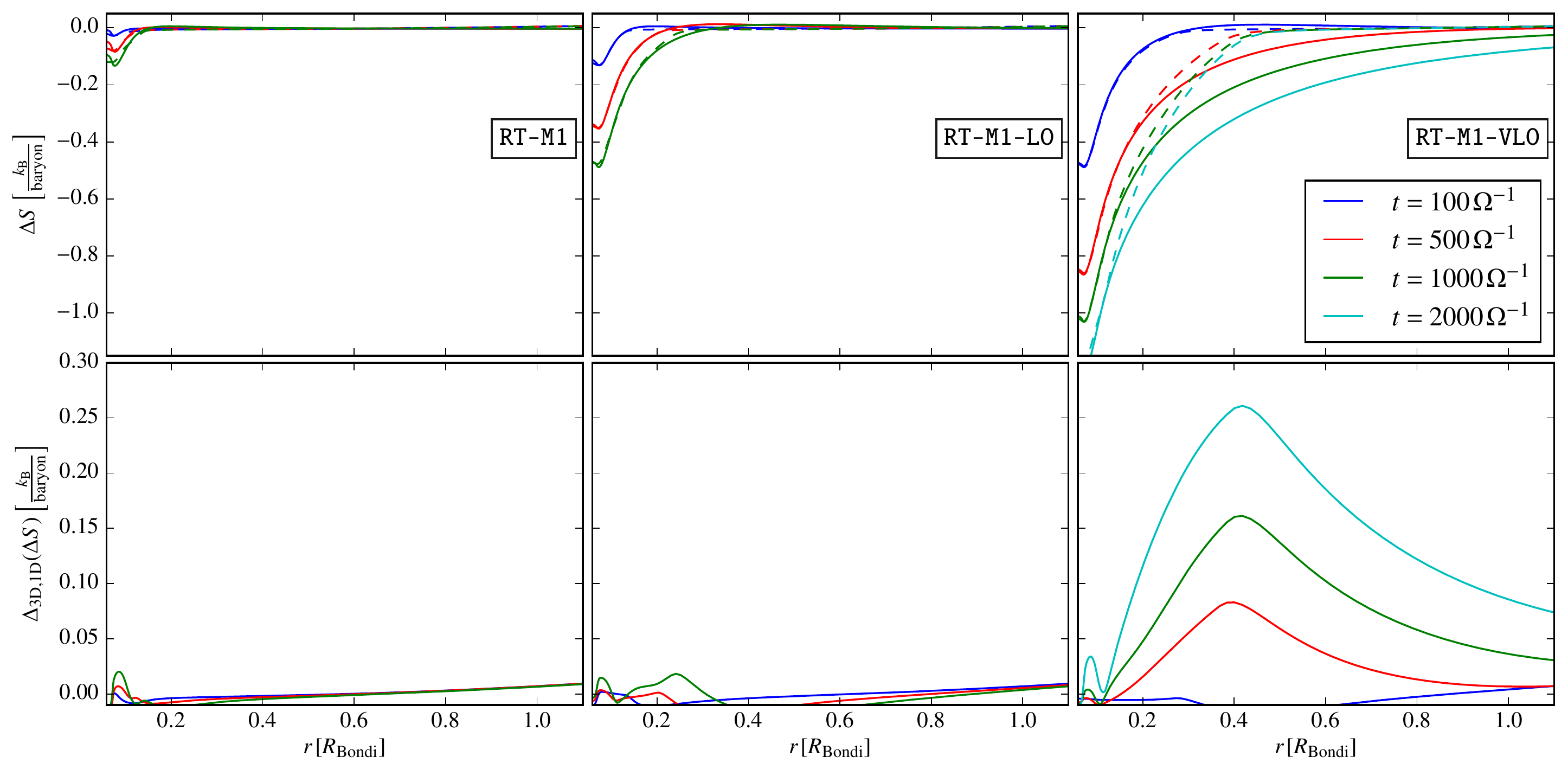}
	\caption{Time evolution of the radial profile of the spherically averaged specific entropy for a $1\,M_\oplus$ core. Colours mark different times while the three columns display results for decreasing opacity from left to right.
	The top row shows both 1D (solid) and 3D (dashed) results while the
	bottom row shows their difference.
	As expected, the radiative cooling is fastest for the lowest optical depth. Quickly,
	the entropy in the 1D atmosphere deviates from the 3D result, which effectively remains at the initial entropy for
	$r \gtrsim 0.4 \, R_\mathrm{B}$. This discrepancy grows as the cooling stalls in the 3D case, while the 1D atmosphere continues
	to lose entropy. To illustrate the convergence to a steady state for the 3D model \texttt{RT-M1-VLO}, the cyan lines in the right panel
	show the state at twice our usual simulation time $t=2 t_\mathrm{end}$. Note that the gradient inversion close to $r=0$ is an artefact of the smoothed gravity.}
	\label{fig:stilder}
\end{figure*}
We next illustrate the effects of recycling on the entropy with a direct comparison to 1D models.
In Fig. \ref{fig:stilder} we show
the spherically averaged radial entropy profile for an Earth-mass core, directly comparing 1D (solid) and 3D (dashed) in the top row and showing their difference $\Delta_{\mathrm{3D,1D}}(\Delta S)$ in the bottom row. Results are given
for different opacities (columns) at different times (colours).
Starting close to the unperturbed mid-plane entropy, the innermost regions that cool the most due to radiation lose entropy, which happens
most efficiently for the lowest opacity.
Initially, 1D and 3D simulations closely follow each other.\footnote{Initially, when radiative cooling has not altered the entropy significantly, 3D simulations can have also slightly lower entropy, seen as negative $\Delta_{\mathrm{3D,1D}}(\Delta S)$. This can be understood since high-$S$ gas can be removed by advection in 3D, which is not possible in 1D.} However, after $t = 500 \,\Omega^{-1}$, we already see a difference emerging for
the \texttt{VLO} runs. While the 1D simulation continues to lose entropy by radiative cooling,
deviating from the initial value all the way out to
$r > R_\mathrm{Bondi}$, the 3D atmosphere effectively maintains its initial entropy in regions $r \gtrsim 0.4 \, R_\mathrm{Bondi}$
due to the combined effect of advection and radiative cooling,
showing that most of the gas in the Bondi sphere is not losing entropy as efficiently as in the 1D case.
In regions where the entropy is higher in 3D compared to 1D ($r \simeq 0.2 - 1 \, R_\mathrm{B}$), the
density is lower as can be seen in Fig. \ref{fig:mid_rho_vphi}. This nicely illustrates how high entropy counteracts contraction.
We find similar results the for higher mass case \texttt{RT-M2-VLO}.

With time this discrepancy between 1D and 3D atmospheres grows mainly due to the fact that the evolution of the 3D atmosphere slows down, suggesting convergence to a quasi-steady state that is close to adiabatic in most of the Bondi sphere.
Clearly, the effects of the low-opacity run are most pronounced, because the radiative diffusion time-scale scales inversely with opacity. But for the intermediate-opacity (\texttt{LO}) run we also see the first signs of a mismatch emerging after $t=10^3\,\Omega^{-1}$. In fact, the results show self-similarity: the $\kappa_\mathrm{R} = 10^{-3}$ and $10^{-4}~\mathrm{cm}^2\, \mathrm{g}^{-1}$ runs in panel (a) match those of $\kappa_\mathrm{R} =10^{-2}$ and $10^{-3} ~\mathrm{cm}^2\, \mathrm{g}^{-1}$ in panel (c). Also note that at 0.1 au, $10^3\,\Omega^{-1}$ only corresponds to a physical time of $5\,\mathrm{yr}$! We therefore expect that the behaviour seen in the \texttt{VLO} run will (i) continue to diverge from the 1D evolution; and (ii) will be representative for the higher opacities.

To illustrate the convergence towards a steady state for simulation \texttt{RT-M1-VLO} we have evolved this simulation for twice as long, up to $t= 2000 \, \Omega^{-1}$. The results are shown as the cyan lines in the right panel of Fig. \ref{fig:stilder}. Clearly, dashed green and cyan lines follow each other closely with deviations only in the deepest layers of the atmospheres, which are not as effectively replenished.

\section{Discussion}
\subsection{Recycling for radiative atmospheres}
Similar to \citetalias{Ormel2015II}, we found that the Bondi and Hill sphere are dynamically interacting with the parent protoplanetary disc by constantly exchanging gas. Accounting for radiative transport processes, we found a flow structure similar to the isothermal 3D atmospheres. Most importantly, there is a continuous supply of high entropy gas from the disc that is brought deep into the atmosphere along horseshoe and shearing orbits. Thus, we confirm that such atmospheres cannot be described as isolated. For the higher mass planets considered in this work, a larger fraction of the Bondi sphere is dynamically dominated by the shearing flow (compare Fig. \ref{fig:M2-VLO-flow} to \citetalias[][Fig. 2]{Ormel2015II}).
Different from \citetalias{Ormel2015II}, however, we do not find out-spiralling streamlines in the mid-plane, but circular streamlines close to the planet (see Sect.~\ref{sec:inner-core}).

Recently, \citet{WangEtal2014} and \citet{FungEtal2015} also reported that isothermal, embedded atmospheres represent open systems. Studying planets with lower thermal masses ($R_\mathrm{Bondi}/H \sim 0.1$), \citet{WangEtal2014} found circumplanetary discs forming.
Their larger orbital distance at $5.2 \,\mathrm{au}$ and the assumed isothermality are most likely beneficial for the formation of such a disc, since we do not find this feature for planets with even higher thermal masses.  In a global simulation using refined grids towards the planet and also solving for radiation transport, \citet{DAngelo2013} identified bound envelopes of sizes $\sim 0.4 - 0.9 \,R_\mathrm{B}$, far larger than our case. Their parameter space is still considerably different from ours as they used a global domain, explicit gas viscosity, non-ideal effects (hydrogen dissociation and ionization) and planetesimal accretion. Therefore, it is still unclear which mechanism is responsible for isolating the pre-planetary atmosphere. But it is not radiation transport.

\subsection{Emergence of inner, low-entropy core}
\label{sec:inner-core}
Although we cannot identify a strictly bound region in any of our simulations, we do find that a low entropy core emerges around the planet and that this region is associated with more circular streamlines (Fig. \ref{fig:Sflow-mid}) and longer replenishment time-scales (Fig. \ref{fig:McumMdot}). This somewhat discrete picture (low $S$, more bound, region interior to $r=0.2 \, R_\mathrm{Bondi}$; high-$S$, thoroughly recycled region outside) nonetheless differs from \citetalias{Ormel2015II}. Apart from the more realistic thermodynamics, it should also be realized that in this work we focused on the long-term behaviour, while sacrificing resolution (which mattered according to \citetalias{Ormel2015II} to be important).  In addition, the adopted gravitational smoothing in our model renders the study of the innermost regions somewhat problematic; a pure Newtonian potential is far preferable (but runs into problems with \texttt{PLUTO}, see \citetalias{Ormel2015II}). Runs at higher resolution
(Fig. \ref{fig:resEntropy}) and different values for $r_\mathrm{smooth}$
(Fig. \ref{fig:rhoflow-mid-Smoothing}) do show a slightly different flow pattern for the inner core region, mainly affecting the rotational feature,
but also that those parameters do not affect the main results. This is discussed further in
the Appendix \ref{app:res}.

A relevant question is how this inner low-$S$ region will evolve on evolutionary time-scales (recall again that the $\sim$$10^3$ orbits only corresponds to 5~yr). From Fig.~\ref{fig:Sflow-mid} one observes that the low-$S$ region expands with time. Possibly, this may imply that the core becomes more isolated, although Fig.~\ref{fig:McumMdot} illustrates that radial recycling rates remain robust.\footnote{Unfortunately, because of the low resolution, we were unable to conduct a topology analysis, that is, it is possible that some of the streamlines in the inner region are closed.} Likewise, the low-opacity run in Fig.~\ref{fig:stilder} also shows that the point where the 3D entropy profile detaches from the 1D curve moves inwards with time: $r=0.19$ at $t=500\,\Omega^{-1}$ to $r=0.14$ at $t=10^3\,\Omega^{-1}$. These are clear manifestations of continued recycling. Finally, physical effects,  like dissociation of $\mathrm{H}_2$ and evaporation of silicate grains or pebbles, may decide the evolution of the inner core (possibly already during the assembly of the super-Earth planet).
From the above results it is clear that the efficiency of recycling strongly depends on the fraction of atmospheric gas that is replenished with gas from the disc. Given the above described topology of the flow, the distribution of mass as a function of distance from the core plays an important role. From Fig. \ref{fig:McumMdot}, we see that in our simulations, where we fixed $\gamma = 7/5$ for simplicity, the mass of the atmosphere is dominated by the outer parts that show significant recycling. However, \citet{PisoYoudin2014} and \citet{Lee2014} argue that close to the planet, where the temperature is sufficiently high ($T \geq 2500\,\mathrm{K}$), accounting for dissocation of $\mathrm{H}_2$ will
result in more realistic values of $\gamma \leq 4/3$. Depending on the ratio of the core radius to the outer radius of the
atmosphere $r_\mathrm{c}/R_\mathrm{B,H}$, this value of $\gamma$ can represent a transition to an atmospheric structure, where most of the
mass accumlates close to the core. If this is the case, recycling time-scales deep in the atmosphere could dramatically increase. However, by integration of the exact hydrostatic density profile (not the power-law approximation) we find that for values of $r_\mathrm{c}/R_\mathrm{B,H}$ (see Table 2) corresponding to our planets, this transition occurs at an adiabatic index of only $\gamma \simeq 1.2$. Still, non-ideal effects as H-dissociation (resulting in non-uniform $\gamma$) could yet render the atmosphere centrally condensed, but this we leave as an exercise to be investigated by future models. \minc{For a more realistic treatment, future studies should also consider tabulated, non-constant opacities.}
Altogether, given the short physical times covered, it is hard to predict the long-term thermal and dynamical evolution of this critical region based on the simulations of this work.

\subsection{Implication: Survival of Super-Earths planets}
But these considerations only apply to the inner core; in most of the super-Earth's atmosphere recycling will be robust, consistent with the results of \citetalias{Ormel2015II}. In \citetalias{Ormel2015II}, we showed that the time-scale for atmospheric recycling is a steep function of disk orbital radius $a$: $t_\mathrm{recycling} \sim a^{2.75}$.\footnote{The exact value of the exponent in this relation depends on the choice of the disc model. This value corresponds to the MMSN, however this does not affect the argument made here.} Physically, the strong dependence on orbital radius follows from the fact that the planet encounters more gas at a higher rate in the inner disk. Recycling will hence be more efficient for close-in orbits, where $t_\mathrm{recycling} \ll t_\mathrm{cool}$, and provides in particular for super-Earth planets an avenue to prevent the premature collapse of their envelopes. With the radiation-transport simulations of this work, the recycling hypothesis, first suggested by \citetalias{Ormel2015II}, has been put on a firmer ground.

It is encouraging that our simulations readily yield envelopes with mass fractions of a few percent, consistent with current predictions from observations \citep[e.g.][]{WolfgangLopez2015}. \minc{On time-scales of a disc lifetime, we expect these values to be subject to change. Further growth or atmospheric loss due to above mentioned mechanisms will affect the envelope. Here, we just argue that recycling does not restrict atmosphere masses to values that are too low, given that super-Earths formed in a gas-rich environment.}

The maximum mass of the planets that we tested has been fixed at $5\,M_\oplus$. At this mass the ratio of the Bondi radius-to-gas scale-height, $R_\mathrm{Bondi}/H=1.9$, which, being larger than unity, suggests that non-linear effects become important. Because the thermal mass reduces with higher temperature discs, we expect that $10\,M_\oplus$ planets in hotter disks will be consistent with the local result of this paper. On the other hand, for a thermal mass much beyond unity we expect that the gravitational backreaction of the disc on the planet atmosphere matters;
effects such as shocks at the spiral arms and gap opening are likely to affect the evolution of gaseous envelopes \citep[e.g.][]{KleyNelson2012}.  While the high resolution around the planet is key to study atmospheric structure, global models will be needed to properly capture the response of the protoplanetary disc in order to study long-term evolution. Approaches such as nested grids \citep[e.g.][]{DAngelo2013} or a coupling of local and global simulations might present a way to achieve this goal in the future for planets in this regime.

\section{Summary}
In this study, we have used hydrodynamical simulations including radiative transport
and directly account for the thermal effects of atmospheric recycling.
Adopting the same spherical geometry in a local shear frame as in \citetalias{Ormel2015I} \& \citetalias{Ormel2015II}, we studied the
evolution and structure of the emerging gaseous envelope for embedded planets with higher
thermal masses, approaching $m \sim 1$.
In particular, we focused on low-mass planets on close-in orbits at 0.1 au, close to where many
super-Earths are found today and added the vertical component of stellar gravity and the resulting disc stratification to our setup.

Our key findings are:
\begin{enumerate}
    \item Allowing for radiation transport, we find that pre-planetary atmospheres represent open systems that show significant rates of recycling, qualitatively and quantitatively (Fig. \ref{fig:McumMdot}) similar to the isothermal runs for very low mass planets of \citetalias{Ormel2015II},with the important exception of the more isolated inner core that we find here.
    \item Recycling equilibrates the entropy between disk and envelope: low-$s$ gas is removed from the envelope, whereas high-$s$ enters the atmosphere, in particular from high latitudes.
    \item Through a comparison of the entropy profile between 1D and 3D geometries, we found that recycling suppresses cooling of the envelope. Already after a few years the 3D flow simulations show significant departure from 1D (hydrodynamically-isolated) atmospheres.
    \item Contrasting the results of isothermal 2D simulations \citepalias{Ormel2015I} but in agreement with 
    isothermal 3D simulations \citepalias{Ormel2015II}, rotation does not play a
    significant role in the emerging atmospheres, which are fully pressure supported.
\end{enumerate}

Accounting for more realistic thermodynamics via radiation transport, we have shown that recycling
constitutes an important mechanism to change the thermal evolution of low-mass planets -- super-Earths of thermal mass $m\lesssim1$ ($M_\mathrm{p} \lesssim G/c_\mathrm{s}^2 H)$ --  compared to isolated 1D atmospheres.
Recycling is therefore an attractive way for super-Earths to survive in gas-rich discs. Staying in the sub-critical regime they can avoid runaway gas accretion and becoming hot Jupiters even when the atmosphere opacities are low.
\section*{Acknowledgements}
We thank many colleagues for helpful discussions that have benefited this work. Also, we would like to thank the anonymous referee for thought- and helpful comments that have substantially improved the quality of this manuscript.
NPC and RK acknowledge financial support via the Emmy Noether Research Group on Accretion Flows and Feedback in Realistic Models of Massive Star Formation funded by the German Research Foundation (DFG) under grant no. KU 2849/3-1. CWO\ is supported by the Netherlands Organization for Scientific Research (NWO; VIDI project 639.042.422). The authors acknowledge support by the High Performance and Cloud Computing Group at the Zentrum f\"ur Datenverarbeitung of the University of T\"ubingen, the state of Baden-W\"urttemberg through bwHPC
and the German Research Foundation (DFG) through grant no INST 37/935-1 FUGG.




\bibliographystyle{mnras}
\bibliography{bbl}




\appendix

\section{Effects of resolution and smoothing on inner core}
\label{app:res}

\begin{figure}
	\centering
	\includegraphics[width=0.48\textwidth]{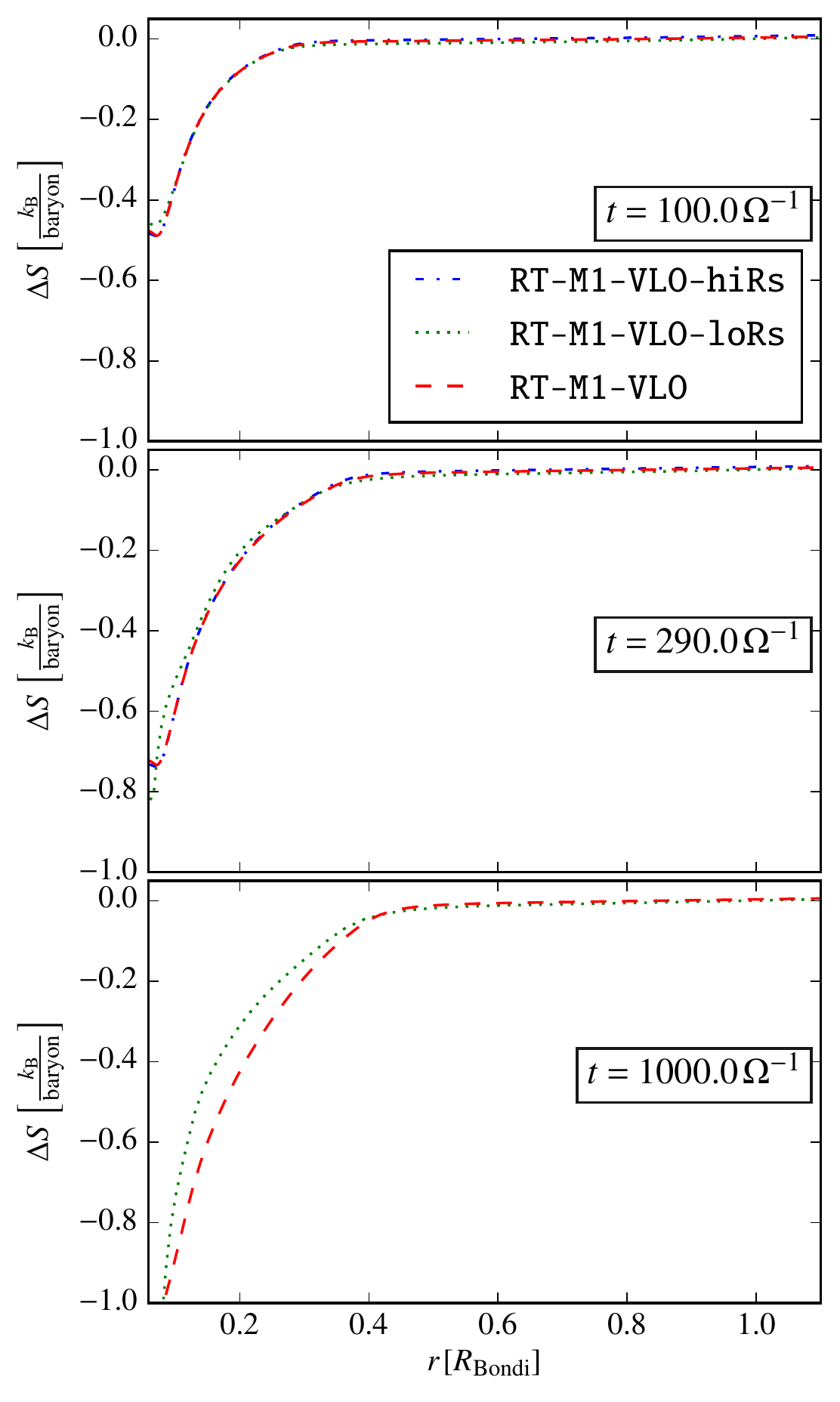}
	\caption{Spherically averaged radial entropy profile for \texttt{RT-M1-VLO} run in different
	resolutions. Only the first two panels include the \texttt{hiRs} results, since this is the
	last available output time for that resolution.}
	\label{fig:resEntropy}
\end{figure}
\begin{figure}
	\centering
	\includegraphics[width=0.48\textwidth]{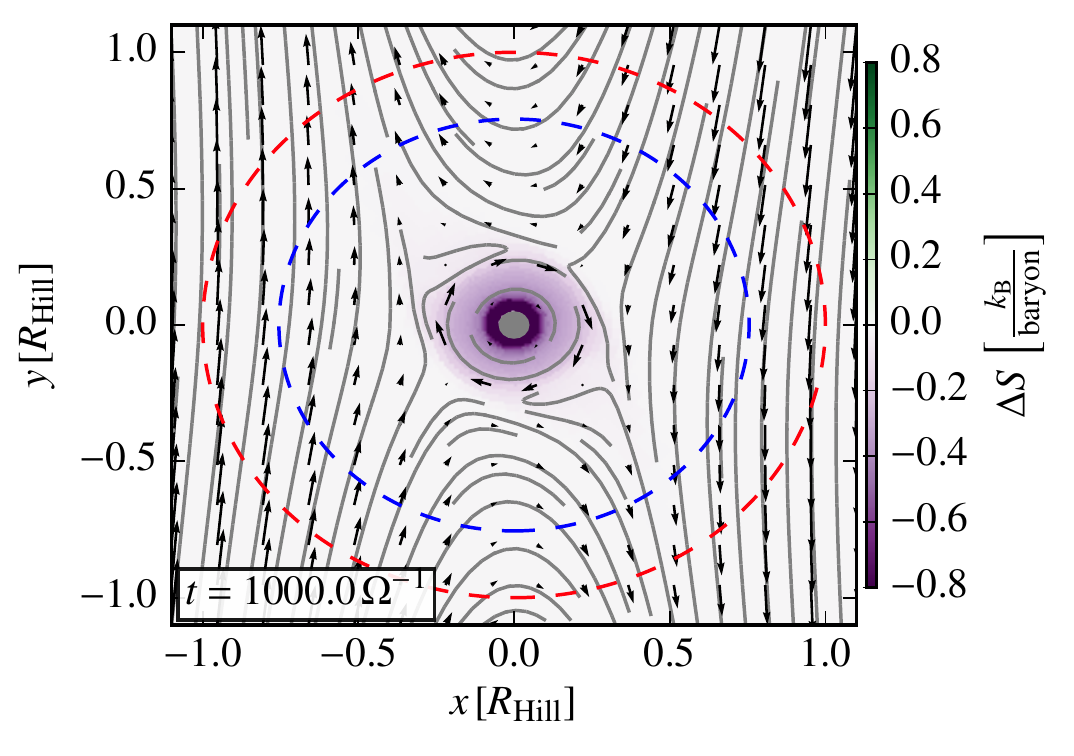}
	\caption{Mid-plane entropy and flow structure for \texttt{RT-M1-VLO-loRs}. Similar to Fig.
	\ref{fig:Sflow-mid}.}
	\label{fig:Sflow-mid-loRs}
\end{figure}
To understand the effects of resolution on the main result of our study, Fig. \ref{fig:resEntropy} shows the spherically averaged entropy profile that was obtained for
three different resolutions. While initially all resolutions agree closely, a small deviation in
the entropy profile is visible at $t=1000 \, \Omega$ between the low and standard resolution.
Mainly above described the inner core is affected, as the \texttt{loRs} maintains slighlty higher
entropy. However, the difference is small and the overall structure remains valid.
Results of \texttt{hiRs} runs agree more closely to those of our standard resolution,
suggesting convergence. Most importantly: Independent of resolution, all runs yield similar values for $r_\mathrm{circ}$, which means that in all runs of the resolutin study, the upper layers are affected by recycling in the same way.
Similar to
the top panel of Fig. \ref{fig:Sflow-mid}, we show the mid-plane structure for the lower resolution run \texttt{RT-M1-VLO-loRs} at the end of the simulation in Fig. \ref{fig:Sflow-mid-loRs} with the same scaling of the color bar. While the overall structure
of the flow is similar, the rotational feature has moved slightly outward and rotation has
decreased. These findings suggests that higher resolution is important for the evolution of the
inner core and such simulations should be conducted also for longer times, once this is
computationally feasible.

\begin{figure}
	\centering
	\includegraphics[width=0.48\textwidth]{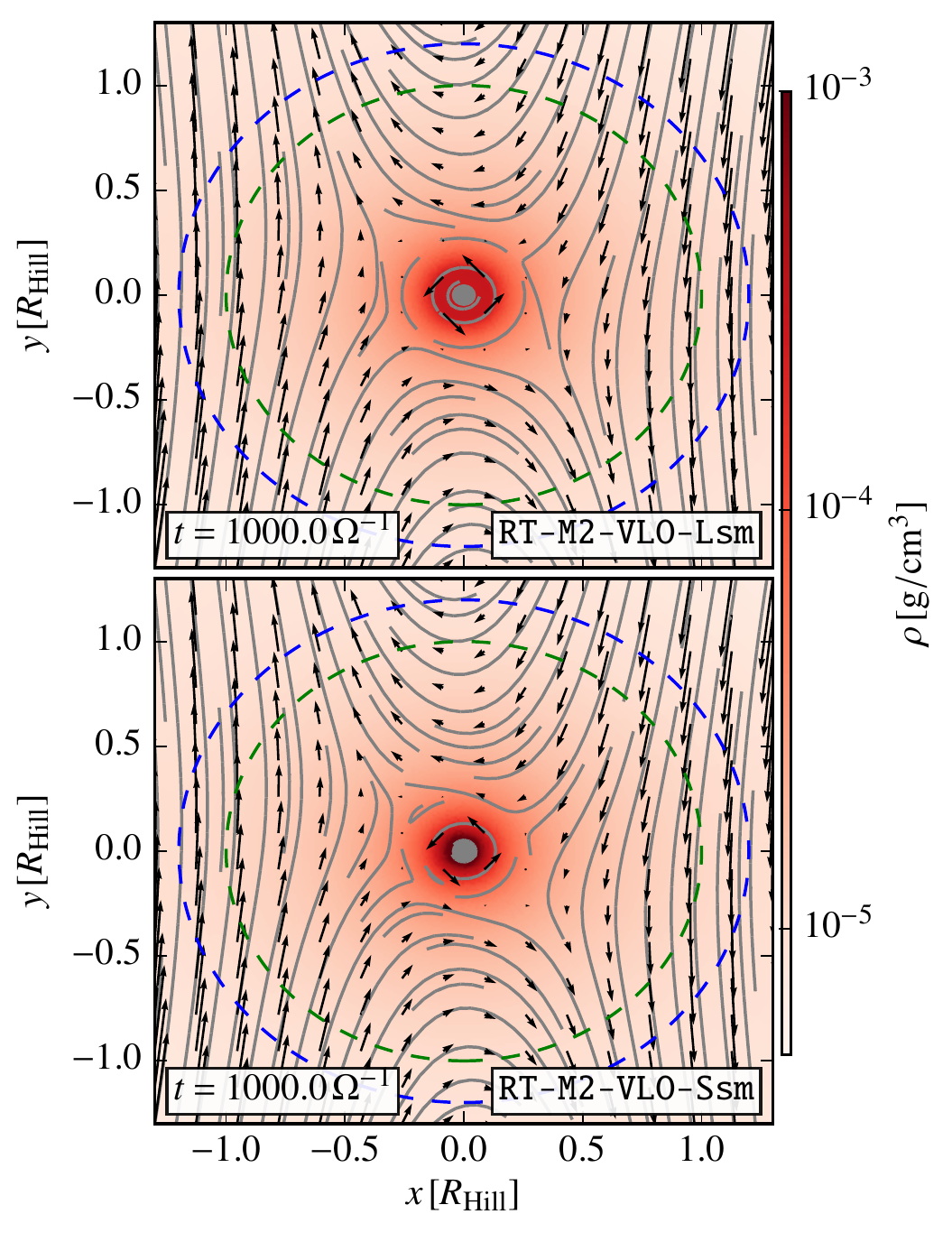}
	\caption{Mid-plane density and flow structure for \texttt{RT-M2-VLO} comparing
	larger (top) and smaller (bottom) values of $r_\mathrm{smooth}$ compared to our standard
	setup.}
	\label{fig:rhoflow-mid-Smoothing}
\end{figure}
For the simulation \texttt{RT-M2-VLO} we conducted test runs with half of the standard smoothing length $r_\mathrm{smooth} = 0.5 \,r_\mathrm{c}$ (\texttt{Ssm}) and twice $r_\mathrm{smooth} = 2 \,r_\mathrm{c}$ (\texttt{Lsm}). A mid-plane view of the flow and densities is shown in Fig. (\ref{fig:rhoflow-mid-Smoothing}).
As expected, the weaker smoothing results in increased densities
close to the core, while stronger smoothing allows for lower densities compared to the
standard case. Similar to our standard smoothing, rotation always stays sub-Keplerian. However,
the sense of the closed streamlines that are embedding the core changes with different
smoothing parameters. For both lower and higher values we find prograde rotation at a similar
distance from the core, contrasting the retrograde feature reported above.
Clearly, further efforts should be made to
implement a true Newtownian potential in future studies of the inner core. We note that
the overall atmosphere structure is not substantially affected by the choice of
$r_\mathrm{smooth}$.


\bsp	
\label{lastpage}
\end{document}